\documentclass{article}
\usepackage[utf8]{inputenc}
\usepackage{graphicx}
\usepackage{geometry}
\usepackage{setspace}
\usepackage{amssymb}
\usepackage{amsmath}
\usepackage{float}
\usepackage{authblk}
\usepackage{ccaption}
\usepackage[style = nature]{biblatex}
\usepackage[dvipsnames]{xcolor}
\usepackage{lineno}
\graphicspath{{Figures/}}
\newcommand{\ket}[1]{| #1 \rangle}
\newcommand{\bra}[1]{\langle #1 |}
\newcommand{\g}{\ket{g}}
\newcommand{\oa}{\ket{ \! \downarrow, 1}}
\newcommand{\ob}{\ket{ \! \uparrow,   1}}
\newcommand{\oc}{\ket{ \! \uparrow,   2}}
\newcommand{\od}{\ket{ \! \downarrow, 2}}
\newcommand{\sq}{{s, n}}
\newcommand{\fr}{f_\mathrm{r}}
\newcommand{\fd}{f_\mathrm{d}}
\newcommand{\Vg}{V_\mathrm{g}}
\newcommand{\vf}{v_\mathrm{F}^s}
\newcommand{\gc}{g_{\mathrm{c}}}

\newcommand{\ts}{\tau_\mathrm{S}}
\newcommand{\ms}{\mathrm{ms}}
\newcommand{\us}{\mu\mathrm{s}}
\newcommand{\ns}{\mathrm{ns}}
\newcommand{\nm}{\mathrm{nm}}
\newcommand{\nA}{\mathrm{nA}}
\newcommand{\um}{\mathrm{\mu m}}
\newcommand{\uT}{\mathrm{\mu T}}
\newcommand{\mT}{\mathrm{mT}}
\newcommand{\mK}{\mathrm{mK}}
\newcommand{\GHz}{\mathrm{GHz}}
\newcommand{\MHz}{\mathrm{MHz}}
\newcommand{\pH}{\mathrm{pH}}
\renewcommand{\thefootnote}{\fnsymbol{footnote}}

\begin{document}

\date{}

\title{Continuous monitoring of a trapped, superconducting spin}

\author[1, *]{M.~Hays}
\author[1, *]{V.~Fatemi}
\author[1]{K.~Serniak}
\author[2]{D.~Bouman}
\author[1]{S.~Diamond}
\author[1,3]{G.~de~Lange}
\author[4]{P.~Krogstrup}
\author[4]{J.~Nygård}
\author[2]{A.~Geresdi}
\author[1, \large{\footnote{Corresponding authors: max.hays@yale.edu, valla.fatemi@yale.edu, michel.devoret@yale.edu}}]{M.~H.~Devoret}

\affil[1]{Department of Applied Physics, Yale University, New Haven, CT 06520, USA}
\affil[2]{QuTech and Kavli Institute of Nanoscience, Delft University of Technology, 2600 GA Delft, The Netherlands}
\affil[3]{Microsoft Quantum Lab Delft, 2600 GA Delft, The Netherlands}
\affil[4]{Center for Quantum Devices, Niels Bohr Institute, University of Copenhagen, 2100 Copenhagen, Denmark}

\maketitle

\renewcommand{\thefootnote}{\arabic{footnote}}

\textbf{
Readout and control of fermionic spins in solid-state systems are key primitives of quantum information processing\supercite{hanson2007spins, childress2013diamond, nakajima2019quantum} and microscopic magnetic sensing \supercite{hong2013nanoscale}.
The highly localized nature of most fermionic spins decouples them from parasitic degrees of freedom, but makes long-range interoperability difficult to achieve.
In light of this challenge, an active effort is underway to integrate fermionic spins with circuit quantum electrodynamics (cQED)\supercite{blais2004cavity, wallraff2004strong, petersson2012circuit, samkharadze2018strong, mi2018coherent, harvey2018coupling, landig2018coherent, mi2018coherent,cubaynes2019highly, borjans2019long, zheng2019rapid, west2019gate, urdampilleta2019}, which was originally developed in the field of superconducting qubits to achieve single-shot, quantum-non-demolition (QND) measurements \supercite{lupacscu2007quantum} and long-range couplings\supercite{majer2007coupling}.
However, single-shot readout of an individual spin with cQED has remained elusive due to the difficulty of coupling a resonator to a particle trapped by a charge-confining potential.
Here we demonstrate the first single-shot, cQED readout of a single spin.
In our novel implementation, the spin is that of an individual superconducting quasiparticle trapped in the Andreev levels of a semiconductor nanowire Josephson element\supercite{Krogstrup15,van2017microwave}.
Due to a spin-orbit interaction inside the nanowire, this ``superconducting spin'' directly determines the flow of supercurrent through the element \supercite{chtchelkatchev2003andreev, padurariu2010theoretical, reynoso2012spin, park2017andreev}.
We harnessed this spin-dependent supercurrent to achieve both a zero-field spin splitting as well as a long-range interaction between the quasiparticle and a superconducting microwave resonator\supercite{tosi2019spin}. 
Owing to the strength of this interaction in our device, measuring the resultant spin-dependent resonator frequency yielded QND spin readout with 92$\%$ fidelity in 1.9 $\boldsymbol{\mu}$s and allowed us to monitor the quasiparticle’s spin in real time.
These results pave the way for new ``fermionic cQED'' devices: superconducting spin qubits operating at zero magnetic field\supercite{chtchelkatchev2003andreev,padurariu2010theoretical, park2017andreev}, devices in which the spin has enhanced governance over the circuit, and time-domain measurements of Majorana modes\supercite{fu2008superconducting, Mourik2012}.
}

Superconducting circuits provide an important set of tools for the creation, manipulation, and measurement of quantum systems.
In cQED\supercite{blais2004cavity, wallraff2004strong}, a quantum system of interest is coupled to a superconducting resonator such that the resonator frequency depends on the system state.
Therefore, by combining superconducting quantum-limited amplifiers\supercite{roy2016introduction} with standard microwave technology, the system state can be non-destructively detected with near-unity single-shot fidelity.
However, integration of this hardware with single quantum spins is made difficult by the inherently weak interaction between the electron magnetic dipole moment and magnetic fields.
It is therefore necessary to couple the spin to the resonator electromagnetic field through an intermediary degree of freedom that interacts with both systems. 
A natural candidate for the mediator is the relativistic coupling between spin and translational degrees of freedom known as spin-orbit interaction. 
This has been used in semiconductor quantum dots to couple the electron spin to a resonator electric field via the electric-dipole moment of the dot charge states\supercite{petersson2012circuit}, but such schemes are constrained by limited dot sizes. 
Here we demonstrate a fundamentally different approach by inductively coupling to the spin-dependent supercurrent of a semiconductor nanowire Josephson element (or Josephson nanowire for short), which has no such limitation.

Similar to conventional quantum dots, a Josephson element composed of a semiconductor nanowire connecting two superconducting reservoirs hosts discrete fermionic modes\supercite{beenakker1991june, Furusaki1991}.
These modes are known as Andreev levels and are occupied by the electronic spin-1/2 quasiparticles of superconductors. 
While a quasi-electron completely confined to a dot cannot participate in charge transport, the quasiparticle occupation of the Andreev levels directly determines the flow of supercurrent through the Josephson nanowire.
Even though the Andreev levels are localized at the junction, the supercurrent can extend over macroscopic distances limited only by the circuit geometry, and thus the Andreev levels can be strongly coupled to a superconducting resonator\supercite{Janvier15, hays2018direct}.
Recently, it has been shown both theoretically\supercite{chtchelkatchev2003andreev, padurariu2010theoretical, reynoso2012spin, park2017andreev} and experimentally\supercite{tosi2019spin} that a Josephson nanowire with an appropriate spin-orbit interaction hosts spin-split Andreev levels and therefore spin-dependent supercurrent.
In this Letter, we combine the fields of confined spins and cQED by detecting the spin state of a quasiparticle trapped in the Andreev levels of a Josephson nanowire.
By inductively coupling the quasiparticle to a superconducting microwave resonator via the spin-dependent supercurrent, we achieve single-shot, QND readout of the quasiparticle spin. 

We first present a qualitative picture of the Andreev levels hosted by Josephson nanowires\supercite{tosi2019spin, van2017microwave, hays2018direct}, which have recently been developed by proximitizing semiconductor nanowires with superconducting contacts.
Andreev levels can be understood as the bound states of a finite square well, with the barriers provided by the superconducting pair potential $\Delta$ of the two superconducting reservoirs [Fig. 1(a)].
Quasi-electrons (quasi-holes) propagating in the nanowire between the reservoirs are Andreev reflected into quasi-holes (quasi-electrons) upon reaching these barriers, a process that conserves the total spin, energy, and approximately momentum but injects a charge of -2$e$ (+2$e$) into the reservoir [Fig. 1(b)].
Localized, spectrally-sharp levels form when the quantum mechanical phases accumulated during a round-trip of propagation and Andreev reflections constructively interfere. 
These levels are usually paired into spin-degenerate doublets, and the number of doublets increases with the both the number of conduction channels and the length of the weak link $\ell$.

In this work, the device was fabricated from an InAs nanowire partially covered in epitaxial Al, with the weak link formed by an $\ell = 500 \; \nm$ uncovered section [Fig. 1(e)]. 
For this $\ell$, the chemical potential in the nanowire can be tuned such that two doublets are present.
In the excitation picture of superconductivity, both doublets are unoccupied in the ground state $\g$ of the Josephson nanowire.
However, superconducting circuits usually exhibit an excess population of quasiparticles that inhabit the continuum of states above the superconducting gap\supercite{aumentado2004}. 
If one such quasiparticle becomes trapped in the sub-gap Andreev levels, its Hilbert space is spanned by the four eigenstates $\ket{\sq}$ of the Hamiltonian $H$.
Here $s = \uparrow, \downarrow$ denotes the quasiparticle spin with the choice of spin label arbitrary, and $n = 1$ or  $2$ labels the lower or higher energy doublet [Fig. 1(a)].
At low temperatures ($\sim20~\mK$), the quasiparticle will reside with high probability in the two spin states of the lower energy doublet. 

Detection of this spin with conventional cQED techniques necessitates lifting the spin degeneracy.
While Kramers theorem does not hold in the presence of a nonzero weak-link phase bias $\varphi$, an additional ingredient is required to split the spin states.
Here this is provided by the spin-orbit interaction present in the multi-subband InAs nanowire.
This interaction causes the quasiparticle spin to hybridize with its translational degrees of freedom and results in an energy-dependent spin texture\supercite{governale2002spin}, though we continue to label these states as $s = \uparrow, \downarrow$ for simplicity.
Critically, this interaction produces a spin-dependent Fermi velocity $\vf$, and therefore a spin-dependent propagation phase, as depicted in Fig. 1(b) for positive momentum. 
The constructive interference condition required for localized levels to form is thereby modified and spin degeneracy is broken, as can be seen from the $\varphi$-dispersion for bound states deep in the gap:

\begin{equation}
    \epsilon(\varphi,s) \cong \pm  \frac{\Delta \hbar \vf/\ell}{2( \Delta + \hbar \vf/\ell)}(\varphi - \pi (2k + 1))
\end{equation}

\noindent where $+/-$ corresponds to positive/negative current-carrying states and $k \in \mathbb{Z}$. 
This relation can be viewed as a competition between two energy scales: the pair potential $\Delta$ and the spin-dependent dwell energy $\hbar \vf/\ell$. 
Such a spin-split spectrum is plotted in Fig. 1(c) before (gray lines, Eq. (1)) and after (colored curves) elastic scattering within the weak link is introduced \supercite{park2017andreev, tosi2019spin}.

While the broken degeneracy is integral to our spin-detection scheme, the higher energy doublet also plays a critical role. 
State readout with cQED relies on the existence of microwave transitions between the states $\ket{m}$ to create a state-dependent dispersive shift $\chi_m$ of the superconducting resonator's frequency. 
The extent to which each microwave transition participates in $\chi_m$ is determined by the coupling operator between the system of interest and the resonator.
Below, we demonstrate that the quasiparticle and the resonator are coupled via an approximately spin-conserving junction current operator $J$. 
As such, neither the direct spin-flipping intra-doublet transition nor the spin-flipping inter-doublet transitions [thin arrows in Fig. 1(c), curves in Fig. 1(d)] contribute appreciably to the dispersive shift. 
The dispersive shifts of the lower doublet states $\chi_{s,1}$ are thus dominated by the two remaining inter-doublet transitions (frequencies $f_s$), which are depicted by the thick arrows in Fig. 1(c) and curves in Fig. 1(d).
Although these transitions are spin-conserving, the shift they induce is nonetheless spin-dependent, which we describe using second-order perturbation theory (see Supplementary Information for details): 

\begin{equation}
    \chi_{s,1} \cong - \frac{\Phi_\mathrm{r}^2}{2\pi \hbar^2} \frac{2 f_{s}}{f_{s}^2 - f_\mathrm{r}^2} |\bra{s, 2} J \ket{s, 1}|^2
\end{equation}

\noindent Here $f_\mathrm{r}=9.188$ GHz is the bare resonator frequency and $\Phi_\mathrm{r}$ is the zero-point fluctuation of the resonator flux drop across the shared inductance [Fig. 1(e)].  
To detect such frequency shifts, we monitored the complex reflection amplitude $\Gamma_{s,1} = I_{s,1} + i Q_{s,1}$ using a microwave readout tone with frequency $f_\mathrm{r}$.
Upon routing the reflected readout tone through a quantum-limited parametric amplifier and integrating for 1.9 $\us$, we found that $\Gamma$ clustered into three distributions [Fig. 2(b)]. 
As we now demonstrate, these distributions can be mapped to $\g$, $\oa$, and $\ob$ based on their dispersive shifts. 

The dispersive shifts $\chi_{s,1}$ and therefore the distribution centers $\Gamma_{s,1}$ can be estimated from the $\varphi$-dependence of the nanowire transition spectrum.
To probe the spectrum, we used an external flux $\Phi$ to set $\varphi \approx 2\pi\frac{\Phi}{\Phi_0} \mathrm{mod}(2\pi)$ and applied a variable frequency drive tone $\fd$ to the nanowire. 
When the tone was resonant with a transition, population was transferred between the Andreev levels, which we detected by measuring shifts in the averaged reflection coefficient $\bar{\Gamma}$ [Fig. 2(c)].
We observed four transitions that we attribute to the inter-doublet transitions based on the qualitative agreement of their $\Phi$-dependence with Fig. 1(d). 
As indicated by the stark contrast in brightness, the drive amplitude required to observe the spin-flipping transitions was at least an order of magnitude larger than was required for the spin-conserving transitions (see Extended Data Fig. 5). 
We attribute this to the drive coupling predominantly via the spin-conserving $J$. 
We fit the spectrum with a simple model in which linearly-dispersing Andreev levels of like spin undergo avoided crossings, e.g. due to elastic scattering (see Supplementary Information for details).
Around $\Phi = 0$, we extracted the slope of the $\oa$/$\ob$ energy splitting $d \Delta \epsilon/d \Phi = 1.8~\nA$.
Together with the device loop area of $2250 ~ \um^2$, this yields a synthetic $g$-factor of the quasiparticle of $\sim 4 \times 10^5$ at low fields. 

Our model of the nanowire spectrum clearly yields $f_s(\Phi)$, but it also allows us to better understand the matrix elements $\bra{s,2} J \ket{s,1}$ needed to calculate $\chi_{s,1}$, which we outline here and detail in the Supplementary Information.
From the fit, we infer a Hamiltonian $H(\Phi)$, and therefore a current operator $J(\Phi) = \frac{d H(\Phi)}{d\Phi}$ over the measured flux and frequency range.
We then fit the $\Phi$-dependent $\chi_{s,1}$ via Eq. (2), yielding the $Q_{s,1}(\Phi)$ plotted in Fig. 2(d).
The only free parameter is $\Phi_\mathrm{r}$, which we find to be within $\sim10\%$ of a calculation based on the circuit parameters.
The qualitative agreement of the model with the measurement indicates that two of the distributions are associated with the states $\ket{s, 1}$.
Moreover, this agreement demonstrates that our crude model of a spin-conserving $J$ describes the quasiparticle/resonator coupling in this regime, although a more sophisticated model will be necessary to understand the complete flux dependence of $\chi_{s,1}$.
The third distribution corresponds to $\g$; all three states are simultaneously visible due to the finite trapping lifetime of a quasiparticle in the nanowire ?junction?, as discussed below. 
While the coupling $\gc \propto |\bra{s, 2} J \ket{s, 1}|$ is spin- and $\Phi$-dependent (see Extended Data Fig. 2), we found a maximum value of $\gc \approx 2 \pi \times $ 35 MHz at $\Phi=\pm0.08 \Phi_0$.

We confirmed our interpretation of the state distributions and transition spectrum by directly measuring the population transfer induced by the microwave drive.
As an example, we present the effect of driving with Gaussian pulses the two transitions available to a quasiparticle initially in $\ob$ [pink dashed arrows in Fig. 3(a, e)].
Two new distributions were revealed, which we attribute to $\oc$ and $\od$ [Fig. 3(b, f)].
Because $\chi_{s,2}$ was approximately described by Eq. (2) but with $f_s \rightarrow -f_s$,  these distributions were located at positive $Q$. 
By varying the amplitude $A$ of the $\ob \leftrightarrow \oc$ pulse, we induced Rabi oscillations of the quasiparticle population between the two doublets [Fig. 3(c)].
To the best of our knowledge, this is the first example of quantum control of an individual quasiparticle excitation of a superconductor. 

We next inspected the relaxation dynamics of the trapped quasiparticle.
We found that after the quasiparticle was transferred from $\ob$ to $\oc$ ($\od$) it decayed preferentially to $\ob$ ($\oa$) [Fig. 3(d), short-time behavior in Fig. 3(g)] within a few microseconds, indicating that the dominant spontaneous relaxation was spin-conserving.
We do not currently understand the mechanisms contributing to this relaxation, and we note that the relaxation timescale of  Purcell decay through the resonator mode should be three orders of magnitude longer (see Supplementary Information).
Following the $\ob \leftrightarrow \od$ pulse, the initial spin-conserving relaxation resulted in an average spin polarization of the quasiparticle in the lower doublet, which then decayed on a timescale $\ts = 42 \pm 2 ~ \us$ [Fig. 3(g)].
Such inter-doublet spin-flipping pulses followed by spin-conserving decay could thus be used to initialize the spin state of a trapped quasiparticle.

The above results demonstrate that a trapped quasiparticle is a coherent object and that it resides with near-unity probability in the two low-energy spin states $\oa$ and $\ob$.
We now proceed to an analysis of the undriven dynamics of the nanowire and our spin-detection fidelity.
We first tuned the flux bias to $\Phi = 0.10 \Phi_0$ to maximize the separation of the $\g$, $\oa$, and $\ob$ distributions. 
For a given $1.9 ~\us$ measurement shot, we determined the system state based on the thresholds indicated by the black dashed lines in Fig. 4(a).
We observed quantum jumps between these states by applying a continuous readout tone and partitioning the reflected signal into consecutive shots [Fig. 4(b)]. 
Similarly to previous reports\supercite{Janvier15,hays2018direct}, we found that a quasiparticle remained trapped in the nanowire weak link for $31 \pm 1~\us$ on average.  
In addition, we were able to measure the spin lifetime, which we found to be $\ts = 51 \pm 4~\us$ at this particular phase bias.
Both types of transitions limited the fidelity of our spin readout.
For perfectly QND measurement, consecutive shots should always yield the same result, which means that transitions should never be observed. 
To compare to this ideal, we histogrammed $Q$ conditioned on the state assignment of the previous shot [Fig. 4(c)]. 
We observed that consecutive shots found the same state with high probability.
We quantify these effects via the spin detection quantum non-demolition metric\supercite{touzard2019gated} $\mathcal{F} = (p_{\downarrow 1, \downarrow 1}+p_{\uparrow 1, \uparrow 1})/2$, where $p_{m,m}$ is the probability that two consecutive shots yield the same state $\ket{m}$. 
Here we report $\mathcal{F} = 92.2 \pm 0.1\%$, which, to the best of our knowledge, is the highest published value for a single electronic spin. 

Although a Zeeman effect was not necessary for our detection scheme, interaction with magnetic fields is a fundamental property of spins. 
We determined the spin lifetime $\ts$ as a function of both $\varphi \cong 2\pi\frac{\Phi}{\Phi_0} \mathrm{mod}(2\pi)$ and a magnetic field $B_\perp$ applied perpendicular to the chip substrate [Fig. 4(d)]. 
At $B_\perp = 0~\uT$, we observed that $\ts$ increased with $|\varphi|$ symmetrically about $\varphi = 0$.  
In particular, $\ts = 42 \pm 2~\us$ at $\varphi /\ 2 \pi = 0.085$ in agreement with the free decay measurement shown in Fig. 3(g).  
This dependence of $\ts$ on $\varphi$ is correlated with the energy splitting between $\oa$ and $\ob$, which goes to 0 at $\varphi = 0$ [Fig. 1(d)].
Applying a positive (negative) $B_\perp$ resulted in a positive (negative) shift of the $\varphi$-dependence, which can be explained by a Zeeman-like shift of the Andreev levels\supercite{reynoso2012spin, tosi2019spin} consistent with the observed spectrum at $B_\perp = 380~\uT$ (see Extended Data Fig. 7). 
This correlation of $\ts$ with the $\oa$/$\ob$ energy splitting could be explained by a splitting-dependent nuclei-induced flip rate\supercite{johnson2005triplet} or an energy-dependent bath spectral density, though surprisingly increasing the cryostat temperature did not affect $\ts$ until the temperature reached $\sim 150~\mK$ (see Extended Data Fig. 8).

In summary, we have demonstrated that the spin of an individual quasiparticle trapped in a Josephson nanowire can be detected by coupling the delocalized spin-dependent supercurrent to a superconducting resonator, and that such a quasiparticle can be coherently manipulated.
Looking forward, the realization of cQED-integrated superconducting spin qubits\supercite{chtchelkatchev2003andreev,padurariu2010theoretical, park2017andreev} requires full coherent control over the quasiparticle spin.
This could be achieved through Raman transitions via the higher energy doublet or by applying a magnetic field ($\sim \; 10 - 100~\mT$) to enable direct, $J$-induced intra-doublet microwave driving\supercite{park2017andreev, tosi2019spin}.
Furthermore, for larger fields ($\sim 1 \; \mathrm{T}$) the nanowire could be tuned to a topological phase\supercite{fu2008superconducting, Mourik2012}, where the techniques presented here would reveal the quasiparticle dynamics of the weak link Majorana mode.
As quasiparticle trapping lifetimes will limit both Majorana-based topological qubits and superconducting spin qubits, our measurement scheme applied to such semiconductor-superconductor heterostructures could provide the detailed understanding of quasiparticle dynamics that is essential for future progress.

\begin{figure}
	\centering
	\includegraphics[width=3.153546in]{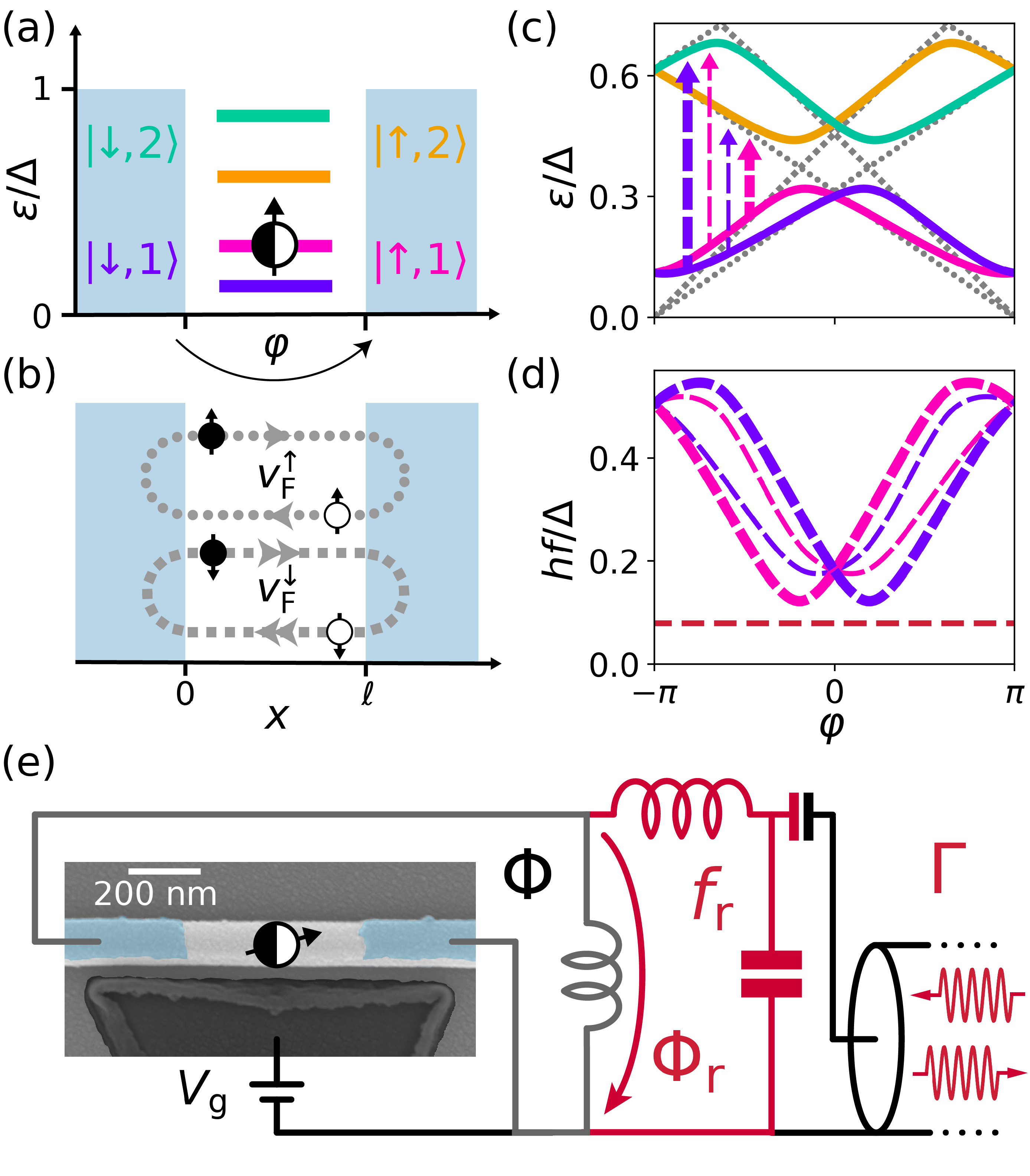} 
	\caption{
    Schematic of spin-orbit split Andreev levels coupled to a microwave resonator. 
    (a) 
    One quasiparticle is trapped in a Josephson nanowire, where its Hilbert space is restricted to two doublets of Andreev levels. 
    (b)
    Due to spin-orbit coupling, charge carriers traverse the weak link with a spin-dependent Fermi velocity $\vf$. 
    Upon reaching a superconducting reservoir, quasi-electrons (black) are Andreev reflected into quasi-holes (white), and vice versa.
    Levels form when these processes constructively interfere. 
    Note that the y-axis is purely diagrammatic and does not represent a physical quantity. 
    (c)
    For a perfectly ballistic channel, the level energies have a linear $\varphi$-dispersion that scales with $\vf$ (gray dotted lines). 
    Crossings at $\varphi = 0$,$\pi$ are protected by time-reversal symmetry, but other crossings are avoided due to elastic scattering in the weak link.
    Transitions out of $\oa$ ($\ob$) are indicated by the purple (pink) dashed arrows, with thin/thick lines denoting whether the spin is flipped/maintained.
    (d)
    Purple and pink curves correspond to dashed arrows in (c) and the maroon line denotes the resonator transition.
    (e)
    Color-enhanced scanning electron micrograph of a Josephson nanowire similar to the measured device (see Methods). 
    The InAs nanowire was partially coated by epitaxial Al (blue), with an uncovered region forming the weak link.
    A flux $\Phi$ applied through a small-inductance loop set the weak-link phase bias $\varphi \cong 2\pi\frac{\Phi}{\Phi_0} \mathrm{mod}(2\pi)$.
    The gate voltage $\Vg$ was used to tune the nanowire such that only two doublets were observed, and was fixed at -1.36 V for all data presented in the main text.
    The Josephson nanowire was inductively coupled to a superconducting resonator (red, frequency $\fr =$ 9.188 GHz), which was capacitively coupled to a transmission line to probe the reflection amplitude $\Gamma$. 
    }
\end{figure}

\begin{figure}
	\centering
	\includegraphics[width= 3.153546in]{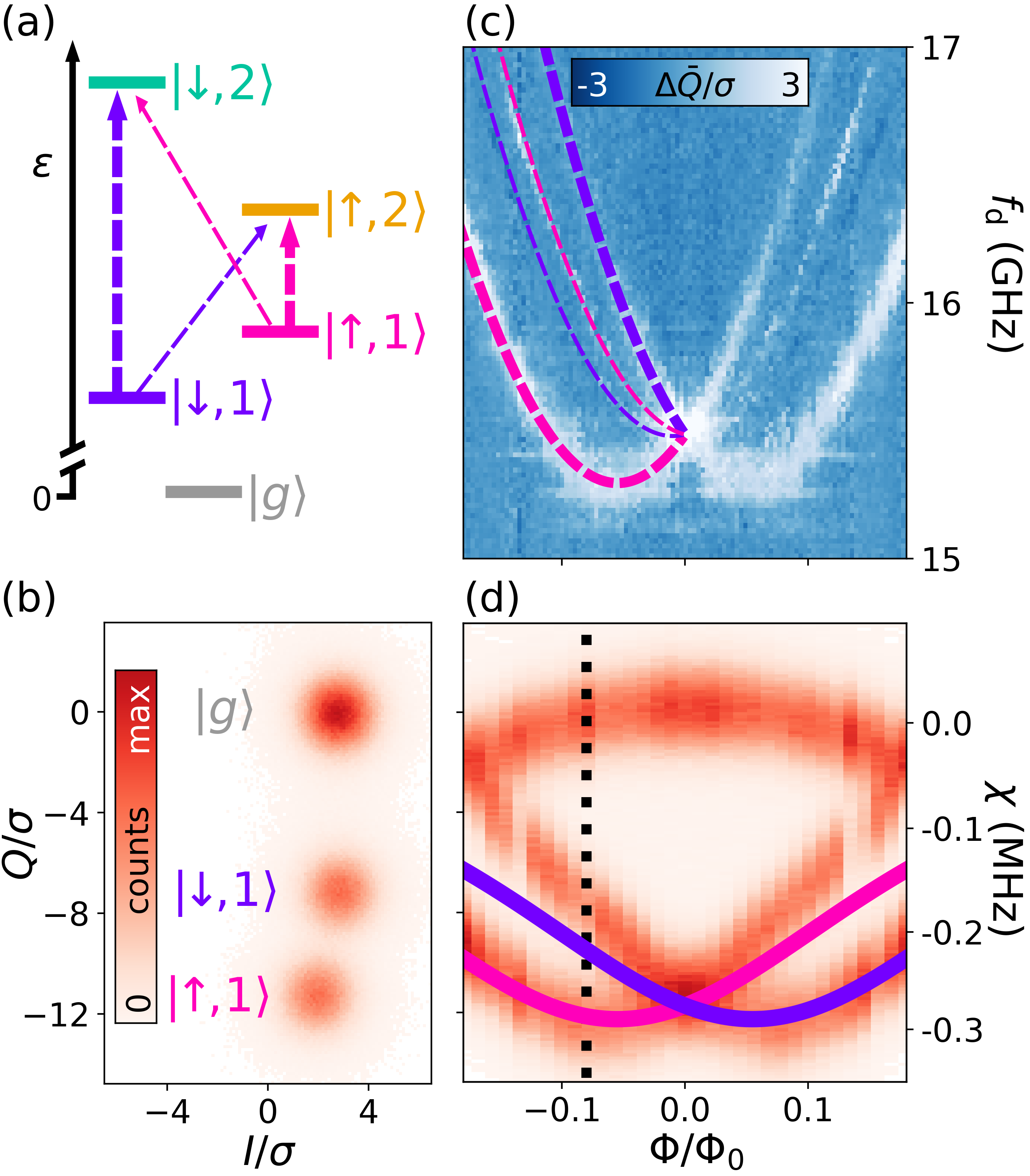} 
	\caption{
	Dispersive readout and spectroscopy of a trapped quasiparticle. 
    (a)
    Level structure and transitions out of the lower doublet for $\Phi < 0$. 
    (b)
    Measured histogram of $\Gamma/\sigma$, where $\sigma$ is the standard deviation of one distribution.
    The data cluster into three distributions, corresponding to $\g$, $\oa$, and $\ob$. 
    (c)
    Drive-probe spectroscopy of the nanowire reveals the four transitions depicted in (a), with fits to a simple model (see Supplementary Materials) plotted for $\Phi < 0$. 
    (d)
    The distributions shown in (b) shift with $\Phi$ as the detuning between the quasiparticle transitions and the resonator varies, from which the absolute dispersive shift (right axis) can be determined. 
    Dashed line indicates $\Phi$ for data in (b), and colored curves are predictions based on the extracted model parameters in (c) with only one additional free parameter (see main text), which captures the scale and shape of the behavior.
    }
\end{figure}

\begin{figure}
	\centering
	\includegraphics[width=6.484248in]{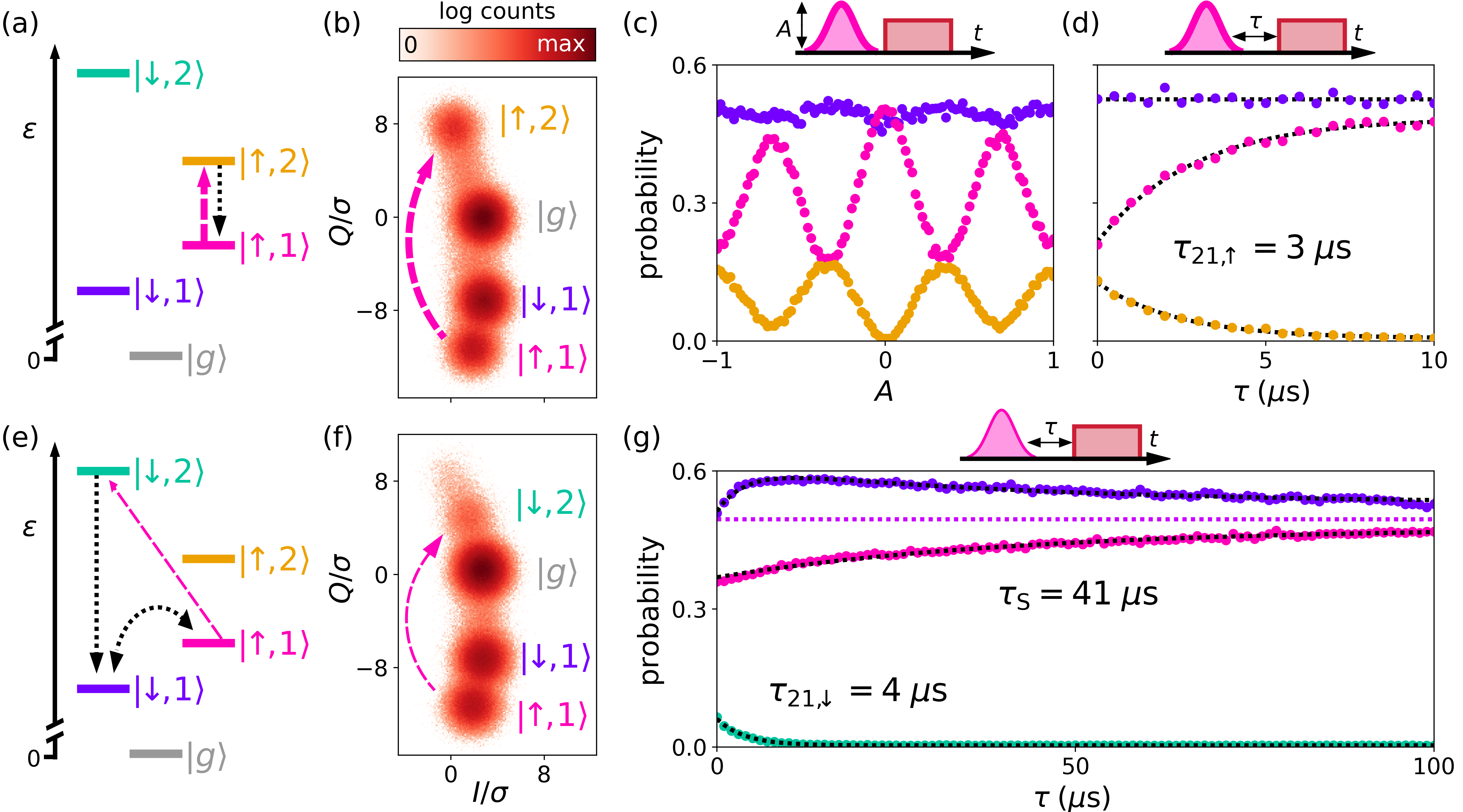} 
	\caption{
	Driven dynamics of a quasiparticle initially in $\ob$ ($\Phi = 0.085 \Phi_0$).
	The quasiparticle was excited into either $\oc$ ((a) through (d)) or $\od$ ((e) through (g)) using Gaussian pulses on the transitions depicted by the pink arrows in (a)/(e). 
	Following each pulse, the distributions corresponding to $\oc/\od$ were visible in the $\Gamma$ histogram ((b)/(f)). 
    In (c), (d) and (g), we plot occupation probabilities for the states of interest as pulse sequence parameters are varied.  
    Probabilities are computed as the number of counts within $2 \sigma$ of the distribution centers, normalized by the steady-state counts.  
    Fits to theory are denoted by dotted black curves (see supplement). 
    (c) Varying the normalized amplitude $A$ of the $\ob \leftrightarrow \oc$ pulse resulted in coherent oscillations of the quasiparticle within the $\uparrow$ manifold. 
    (d) 
    Varying the delay $\tau$ between the $\ob \leftrightarrow \oc$ pulse and the readout pulse revealed exponential decay of the quasiparticle back to $\ob$ with timescale $\tau_{21, \uparrow}$ (black arrow in (a)). 
    (g) 
    Following a $\ob \leftrightarrow \od$ pulse, an initial exponential decay to $\oa$ with timescale $\tau_{21, \downarrow}$ (single-headed black arrow in (e)) resulted in equal and opposite deviation of the $\oa$ and $\ob$ populations from their equilibrium value (magenta dotted line). 
    This spin polarization then exponentially decayed with timescale $\ts$ (double-headed black arrow in (e)). 
}
\end{figure}

\begin{figure}
	\centering
	\includegraphics[width=6.484248in]{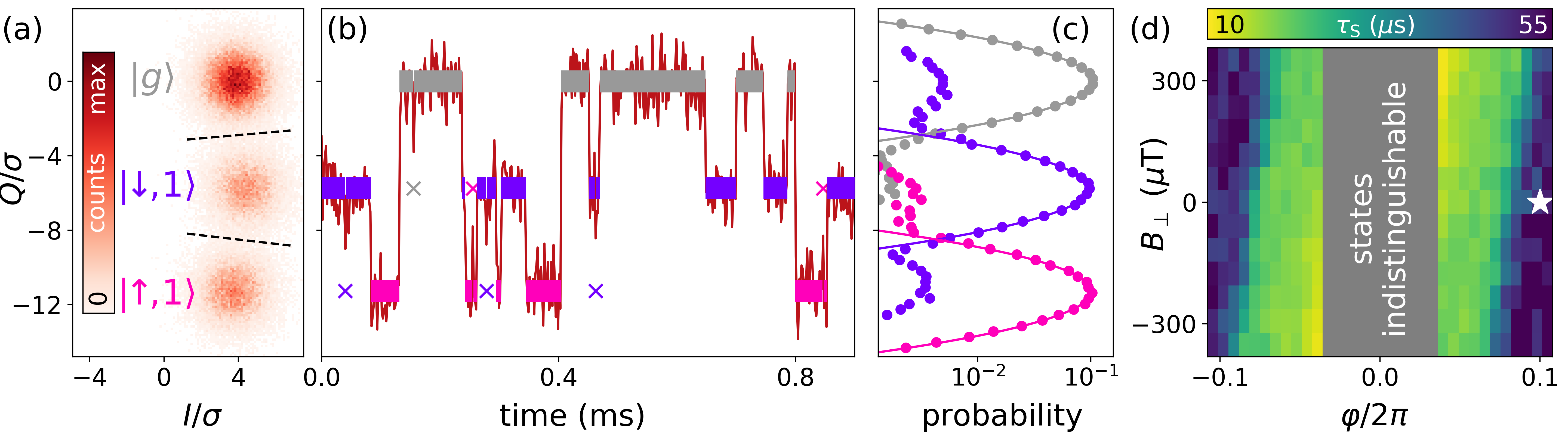} 
	\caption{
    Quantum non-demolition readout of the quasiparticle spin. 
    (a) The system state was assigned to be $\g$, $\oa$, or $\ob$ based on thresholds indicated by the black dashed lines. 
    (b) $Q(t)$ reveals quantum jumps between the three states. 
    Colored bars indicate state assignments, with isolated points indicated by crosses and colored by the most likely state. 
    (c) Histogram of $Q$ conditioned on the state assignment of the previous measurement (indicated by color). Solid lines are gaussian fits.
    (d) By analyzing $\Gamma(t)$ using a hidden Markov model (see Supplementary Materials), the spin lifetime $\ts$ was determined as a function of both $\varphi$ and a magnetic field $B_\perp$ applied perpendicular to the chip substrate. 
    The star indicates the bias for data in (a), (b), and (c).
}
\end{figure}

\printbibliography[keyword={maintext},title={References}]

\section*{Methods}

\textbf{Experiment device and setup.} 
At the time of writing, the device was still inaccessible due to ongoing measurements, so the micrograph displayed in Fig. 1(c) is of a similar $\ell = 500~\nm$ device.
Further device images are provided in Extended Data Fig. 1., along with a full schematic of the cryogenic setup. 
Our device was fabricated on a sapphire substrate. 
After performing microwave simulations of the circuit using Sonnet Suites$^\mathrm{TM}$, we patterned the readout resonator and control structures by electron-beam lithography and reactive ion etching of sputtered NbTiN.
The NbTiN film had a thickness of $150~\nm$ and a sheet kinetic inductance of $0.6~\pH/\mathrm{square}$, which we took into account when calculating the shared inductance between the nanowire and the resonator. 
An MBE-grown [001] wurtzite InAs nanowire with epitaxial Al coating two of six facets was then deposited using a micromanipulator. 
The weak link was defined by selectively wet-etching a $500~\nm$ long section of the Al shell, and contacted to the rest of the circuit using NbTiN. 
After connecting the device to external circuitry [Extended Data Fig. 1.], we cooled it down in a dilution refrigerator with a base temperature of $\sim 20~\mK$. 
We used a coil external to the device to apply a magnetic field approximately perpendicular to the device substrate, which generated the flux $\Phi$. 
The data displayed in Figs. 2, 3 and 4(a-c) were taken at $|\Phi/\Phi_0| < 1$, and as such we interpreted the flux as a phase bias $\varphi \approx 2 \pi \Phi /\Phi_0$. 
The data displayed in Fig. 4(d) was taken using the same coil, but $\Phi$ was swept over approximately 1000 $\Phi_0$. 
For this measurement, we thus interpreted $\Phi$ as both a phase bias $\varphi \approx 2\pi\frac{\Phi}{\Phi_0} \mathrm{mod}(2\pi)$ and a magnetic field $B_\perp = \Phi/A_\mathrm{loop}$. 

\noindent \textbf{Measurement.}
We performed microwave reflectometry of the resonator using a readout tone at the bare resonator frequency of $9.188~\GHz$, which produced an average of $\sim 10$ photons in the resonator during measurement. 
After interacting with the device [Fig. 1(c)], the readout tone was routed through an amplification chain consisting of a SNAIL parametric amplifier\supercite{frattini2018} at base temperature, a HEMT amplifier at 4 K, and finally room temperature amplifiers [Extended Data Fig. 1].
The signal was then down-converted to $50~\MHz$ before being fed into a data acquisition card.
The reflection amplitude $\Gamma$ was computed by comparing this $50~\MHz$ signal to a $50~\MHz$ reference and integrating for $1.9~\us$.

At low gate voltages ($\Vg < -2~\mathrm{V}$), we observed no dependence of $\Gamma$ on the current through our $\Phi$-bias coil. 
As we increased $\Vg$, we observed ranges of $\Vg$ in which $\Gamma$ depended strongly on $\Phi$. 
We attribute this to the transparency of the nanowire conductance channels fluctuating close to one\supercite{van2017microwave, Goffman, hays2018direct}.
To locate the transitions causing these shifts, we performed pulsed drive-probe spectroscopy ($2.5~\us$ drive pulse, $1.9~\us$ readout pulse). 
In the vicinity of $\Vg = -1.36~\mathrm{V}$, we observed the transitions discussed in the main text [Fig. 2(c)]. 
To minimize electric-field-induced decoherence, we made fine adjustments to $\Vg$ such that at $\Phi = 0$ the transitions were at a local maximum in $\Vg$ [Extended Data Fig. 4].
In addition to $\Vg$, we used two additional gates on the proximitized\supercite{chang2015hard} sections of the nanowire to gain additional electrostatic control [Extended Data Fig. 1]. 
Both gates were biased to the same voltage $V_\mathrm{nw} = 0.9$ V for all presented data. 

\noindent \textbf{Definition of synthetic $g$-factor.} Here we define the $g$-factor via the slope of the linear energy splitting between the two spin states under the application of magnetic field: $g = \frac{1}{\mu_\textrm{B}}\frac{d\Delta\epsilon}{d B_\perp}\Bigr\rvert_{B_\perp = 0}= \frac{A_\mathrm{loop}}{\mu_\textrm{B}}\frac{d\Delta\epsilon}{d \Phi}\Bigr\rvert_{\Phi = 0}$.  
Because this $g$-factor depends on the circuit geometry, we dub it ``synthetic''. 

\noindent \textbf{Analysis of driven dynamics.}
The Gaussian pulses used in the experiments depicted in Fig. 3(b-d) had standard deviations of $20~\ns$, while the pulses used in the experiments depicted in Fig. 3(f,g) had $250~\ns$ standard deviations due to the larger total energy required to induce spin-flipping transitions. 
To compute the probabilities plotted in Fig. 3(c,d,g), we first counted the number of shots within 2$\sigma$ of the distribution centers. 
Shots outside of these regions were left unassigned.
Extended Data Fig. 6 illustrates this for the measurement depicted in Fig. 3(c), additionally including counts assigned to the $\g$ population as well as the unassigned counts.
For Fig. 3(c,d,g), we then normalized by the steady-state (undriven) counts for the primary states of interest ($\oa$, $\ob$ and $\oc$ for the Fig. 3(c) measurement).
Due to decay from $\oc$ to $\ob$ during measurement, some shots were mistakenly assigned to $\g$ and $\oa$ or were unassigned because their mid-flight capture resulted in a value of $\Gamma$ that was not associated with any one state distribution. 
This resulted in small oscillations in the apparent populations of these states, large oscillations of the number of shots not assigned to any state (Extended Data Fig. 6), and also what appears to be an unequal probability change between states $\oc$ and $\ob$ in Fig. 3(c). 
The magnitude of these unintended oscillations decreased with shorter integration time, which is consistent with our interpretation; however, the discrimination power also suffered. 
Such decay during measurement also explains the observed $\od$ population in Fig. 3(f), as well as the unequal population deviations at $\tau = 0$ observed in Fig. 3(d,g). 

\noindent \textbf{Quantum jump analysis.}
The spin lifetime $\ts$ and quasiparticle trapping lifetime were extracted from $\Gamma(t)$ using a hidden Markov model algorithm~\supercite{press2015, Janvier15, hays2018direct}.
This analysis assumes that the system possesses three states ($\g$, $\oa$, and $\ob$), and that each state $\ket{m}$ emits values of $\Gamma$ with different (but potentially overlapping) probability distributions $p(\Gamma | m)$. 
Importantly, $p(\Gamma  | m)$ does not need to be known \textit{a priori}. 
By analyzing $\Gamma(t)$, the algorithm yields the most probable $p(\Gamma | m)$, state assignments at each $t$, and transition rates $\gamma_{n,m}$ from $\ket{m}$ to $\ket{n}$.
We measured all six $\gamma_{n,m}$ as a function of $\varphi$, $B_\perp$, and the temperature of the mixing chamber [Extended Data Fig. 6, 8, discussion in Supplementary Information]. 
The spin lifetime was computed as $\ts = 1/(\gamma_{{\uparrow_1}, {\downarrow_1}} + \gamma_{{\downarrow_1}, {\uparrow_1}})$, and the trapping lifetime was computed as $1/(\gamma_{0, {\uparrow_1}} + \gamma_{ 0,{\downarrow_1}})$. 
Note that here we distinguish between the trapping lifetime and the parity lifetime $1/(\gamma_{{\uparrow_1}, 0} + \gamma_{{\downarrow_1}, 0} + \gamma_{0, {\uparrow_1}} + \gamma_{0, {\uparrow_1}}) = 21 \pm 1~\us$, since it is the trapping lifetime which limits the fidelity of the spin detection. 

\printbibliography[keyword={methods},title={References (Methods)}]

\noindent \textbf{Acknowledgements}

\noindent 
We thank Nick Frattini and Vladimir Sivak for providing the SNAIL parametric amplifier.
We are grateful to Marcelo Goffman, Hugues Pothier, Leandro Tosi, and Cristián Urbina for sharing their experimental results and thoughts. 
We acknowledge useful discussions with Sergey Frolov, Luigi Frunzio, Leonid Glazman, Manuel Houzet, Shyam Shankar, Steven Touzard, and Charles Marcus. 

This research was supported by the US Office of Naval Research (N00014-16-1-2270) and by the US Army Research Office (W911NF-18-1-0020, W911NF-18-1-0212 and W911NF-16-1-0349).  
J.N. acknowledges support from the Danish National Research Foundation. 
G.d.L. acknowledges support from the European Union’s Horizon 2020 research and innovation programme under the Marie Skłodowska-Curie grant agreement No. 656129.
D.B. acknowledges support by Netherlands Organisation for Scientific
Research (NWO) and Microsoft Corporation Station Q.
Some of the authors acknowledge the European Union’s Horizon 2020 
research and innovation programme for financial support: 
A.G received funding from the European Research Council, grant no. 804988 (SiMS), and A.G and J.N. further acknowledge grant no. 828948 (AndQC) and QuantERA 
project no. 127900 (SuperTOP).

\noindent \textbf{Contributions}

\noindent M.H., K.S., D.B., G.d.L., A.G., and M.D. designed the experiment.
P.K. and J.N. developed the nanowire materials.
D.B. and A.G. fabricated the device.
M.H. and V.F. performed the measurements.
M.H., V.F., K.S., S.D., and M.D. analyzed the data.
M.H., V.F. and M.D. wrote the manuscript with feedback from all authors. 

\setcounter{figure}{1} 
\renewcommand{\figurename}{Extended Data Figure}

\section*{Extended Data}

\begin{figure}[H]
	\centering
	\includegraphics[width=5in]{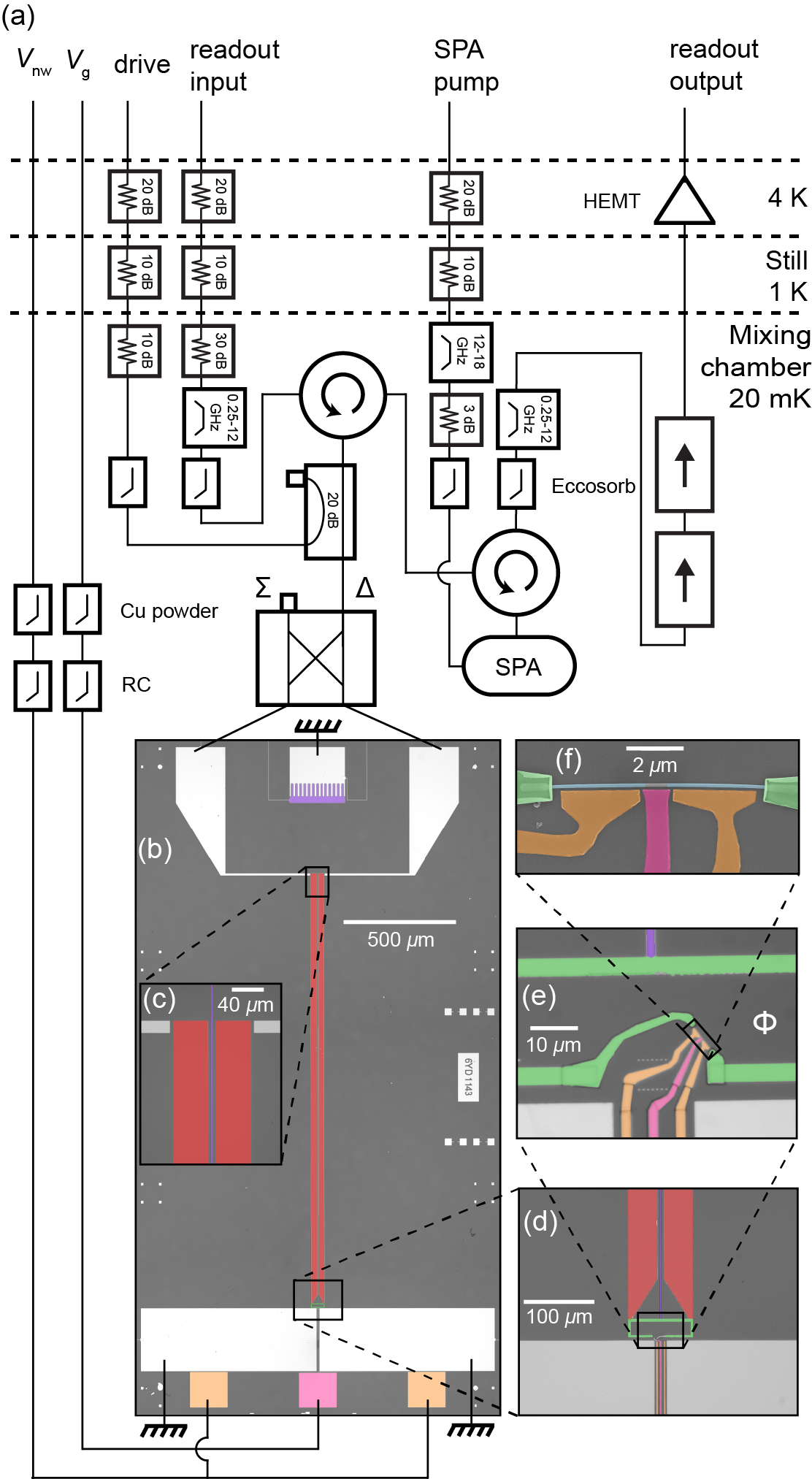} 
\end{figure}

\begin{figure}[H]
  \contcaption{Cryogenic wiring diagram and device micrographs.
    Optical micrograph (e) is of the device on which the presented measurements were performed. 
    Optical micrographs (b), (c), (d) and scanning electron micrograph (f) are of an extremely similar (unmeasured) device, the main difference being that the length of the weak length is $750~\nm$ instead of $500~\nm$.
    The microwave readout and drive tones pass through the depicted circuitry (a) before being routed through the $\Delta$ port of a $180^{\circ}$ hybrid resulting in differential microwave voltages at the device input. 
    After reaching two coupling capacitors (c), the readout tone was reflected off the differential $\sim\lambda/4$ mode of the coplanar strip resonator (red, frequency $\fr = 9.18843~\GHz$, coupling $\kappa_\mathrm{c} = 2\pi \times 1.23~\MHz$, internal loss $\kappa_\mathrm{i} = 2\pi \times  1.00~\MHz$) and then routed through the depicted amplification chain (a), which was comprised of a SNAIL parametric amplifier (SPA), HEMT, and room-temperature amplifiers.  
    In this circuit, the drive tone creates an ac phase drop across the nanowire (f), which is embedded in the superconducting $\Phi$-bias loop (green) at the end of the resonator (d,e).
    One edge of the loop connects the two strips of the resonator and thereby forms the shared inductance with the nanowire.
    We controlled the electrostatic potential in the nanowire weak link (f) with a dc gate (pink, voltage $\Vg$).
    Gates on the nanowire leads (orange) were used to gain additional electrostatic control, which were biased to the same voltage $V_\mathrm{nw} = 0.9~\mathrm{V}$ for all presented data.  
    To reference the resonator/nanowire island to ground, an additional strip runs between the resonator strips, and connects to a large finger capacitor (purple). 
    This strip does not significantly perturb the resonator's microwave properties because it resides at the zero voltage point with respect to the resonator's differential mode.}
\end{figure}

\begin{figure}[H]
	\centering
	\includegraphics[width=3in]{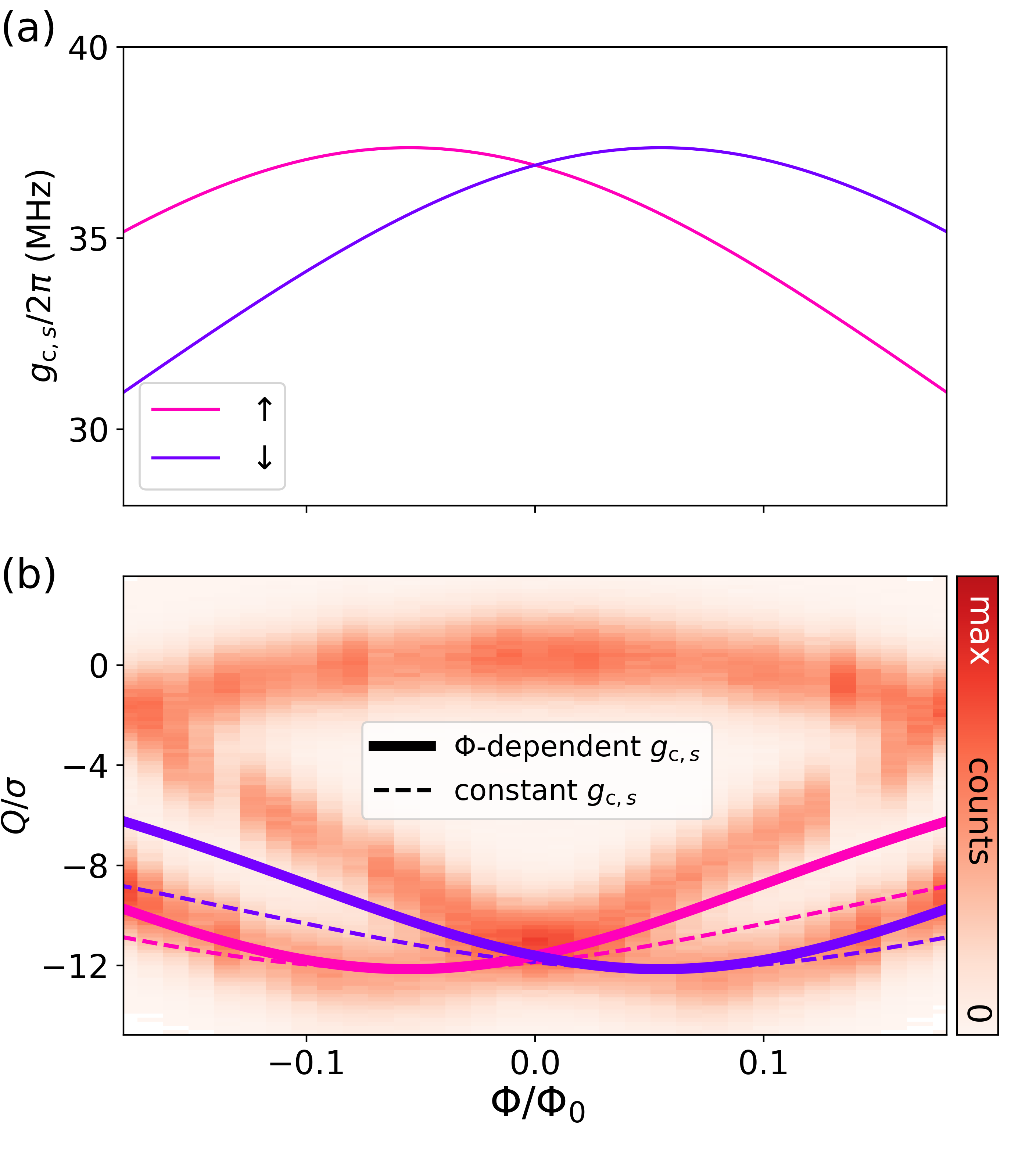} 
	\caption{
    (a) Extracted coupling strengths $g_{\mathrm{c}, s}$ for the two inter-doublet spin conserving transitions. 
    The peaks coincide with the minimum frequency of the transitions ($\Phi = \pm\Phi_
    \mathrm{cross}$) because this is where the mixing between current and energy eigenstates is strongest (see Supplementary Information). 
    (b) Same data as shown in Fig. 2(d).
    Solid lines are the predicted $\chi_{s,1}$ as in the main text, and dashed lines are the $\chi_{s,1}$ if $g_{\mathrm{c}, s}$ is assumed to be constant at its maximum value. 
    }
\end{figure}

\begin{figure}[H]
	\centering
	\includegraphics[width=3in]{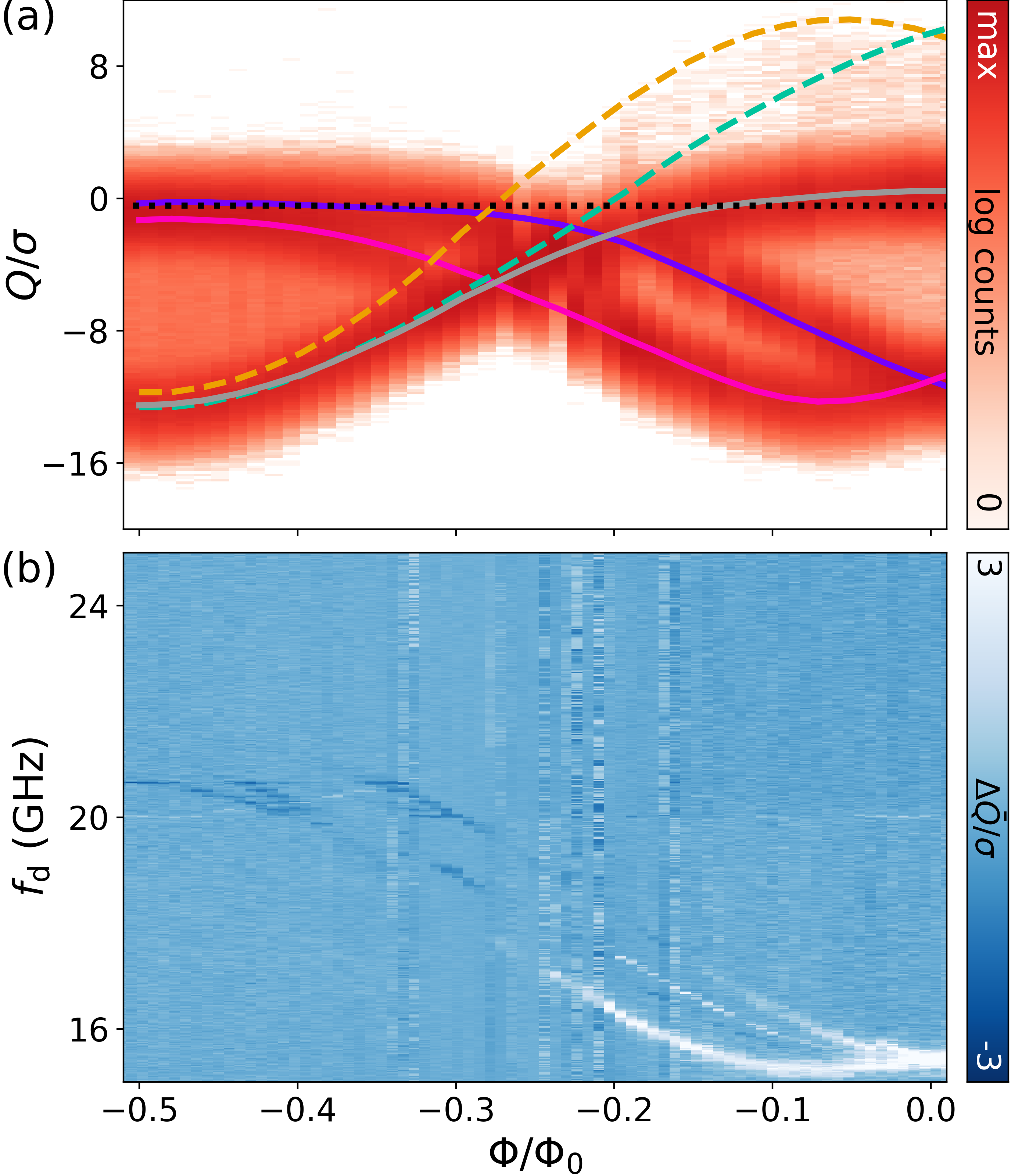} 
	\caption{(a) $\Phi$-dependence of Q over a full half flux quantum.
	The $\oa$ and $\ob$ distributions (traced with purple and pink splines respectively) remain below the bare resonator $Q$ (black dotted line) over the full $\Phi$ range, indicating negative dispersive shifts which are inconsistent with $\chi$ resulting from coupling to the inverse inductance operator.
	The dispersive shift of $\g$ (traced with the gray spline) is likely due to a pair transition with frequency above our measurement bandwidth.
	We also observe a small number of counts around $\Phi = 0$ at positive $Q$, indicating a residual quasiparticle population in $\oc$ and $\od$.
    Assuming the observed dispersive shift of $\g$ is due only to the properties of the lower doublet, the dispersive shift of a quasiparticle in the upper doublet should be given by $\chi_{s,2} = -\chi_{s,1} + \chi_0$. 
    Based on this formula and the plotted splines, we estimated the $\Phi$-dependence of the $\od$ and $\oc$ distributions (dashed, teal, and yellow).
    The predictions track roughly with the residual counts in the vicinity of $\Phi = 0$ before crossing the bare resonator Q. 
    (b) 
    Spectroscopy over the same flux range. 
    The inter-doublet transitions have maximum frequency at $\Phi = -0.5 \Phi_0$, consistent with Fig. 1(e). 
    We attribute the sign change in the measured $\Delta \bar{Q}$ to the crossings of $\chi_{s,1}$ with $\chi_{s_2}$ indicated in (a).
    }
\end{figure}

\begin{figure}[H]
	\centering
	\includegraphics[width=3in]{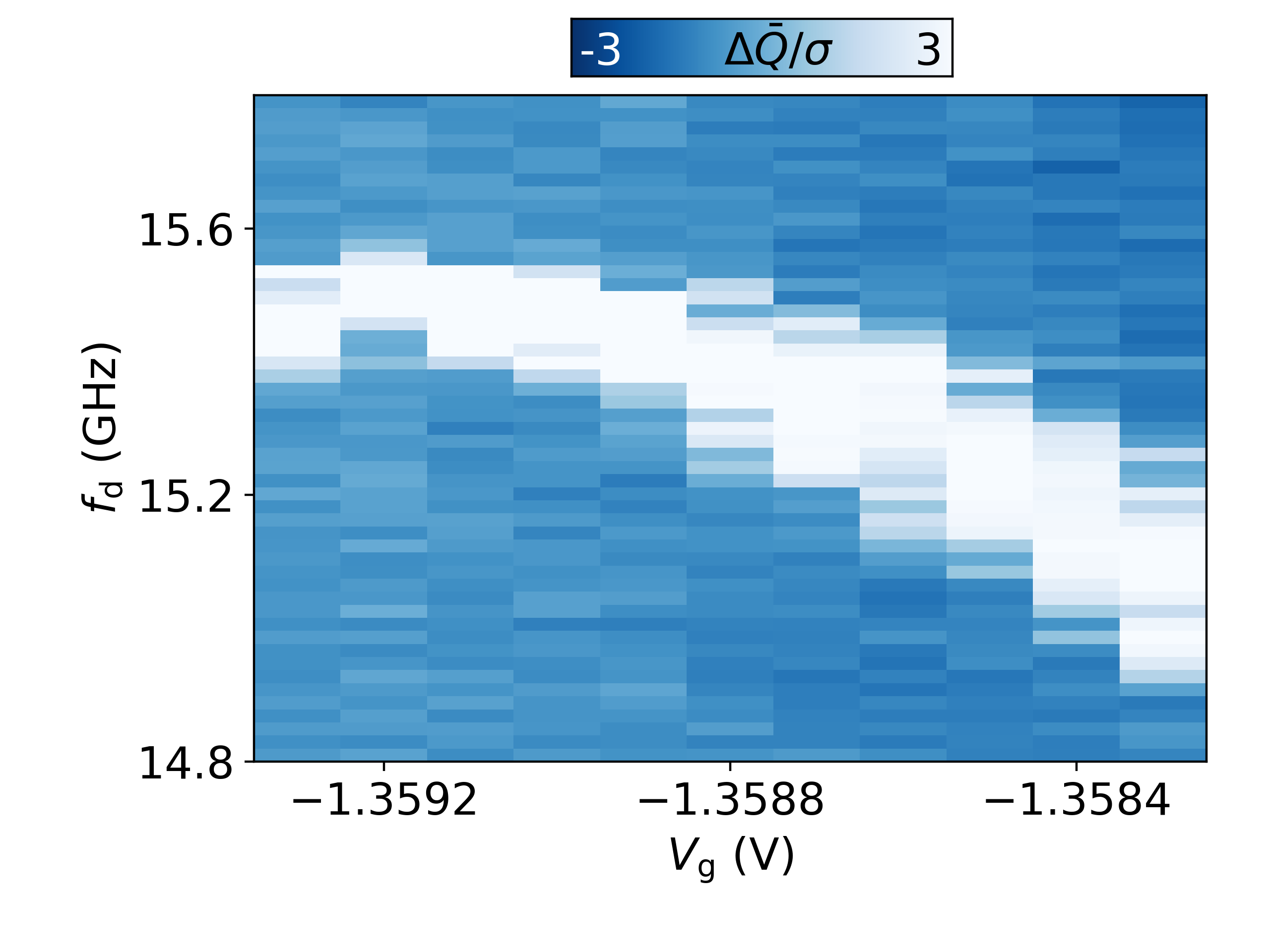} 
	\caption{
    Spectroscopy of the inter-doublet transitions at the $\Phi = 0$ degeneracy point while $\Vg$ is varied. 
    The transition frequency changes due to mesoscopic conductance fluctuations\supercite{van2017microwave, Goffman, hays2018direct}, and a local maximum is observed around $\Vg = -1.3592~\mathrm{V}$.
    The linewidth is visibly narrower at this local maximum, indicating that electric field noise is the dominant source of dephasing. 
    To minimize this dephasing, we performed the measurements presented in the main text at $\Vg = -1.3592~\mathrm{V}$. 
    }
\end{figure}

\begin{figure}[H]
	\centering
	\includegraphics[width=3in]{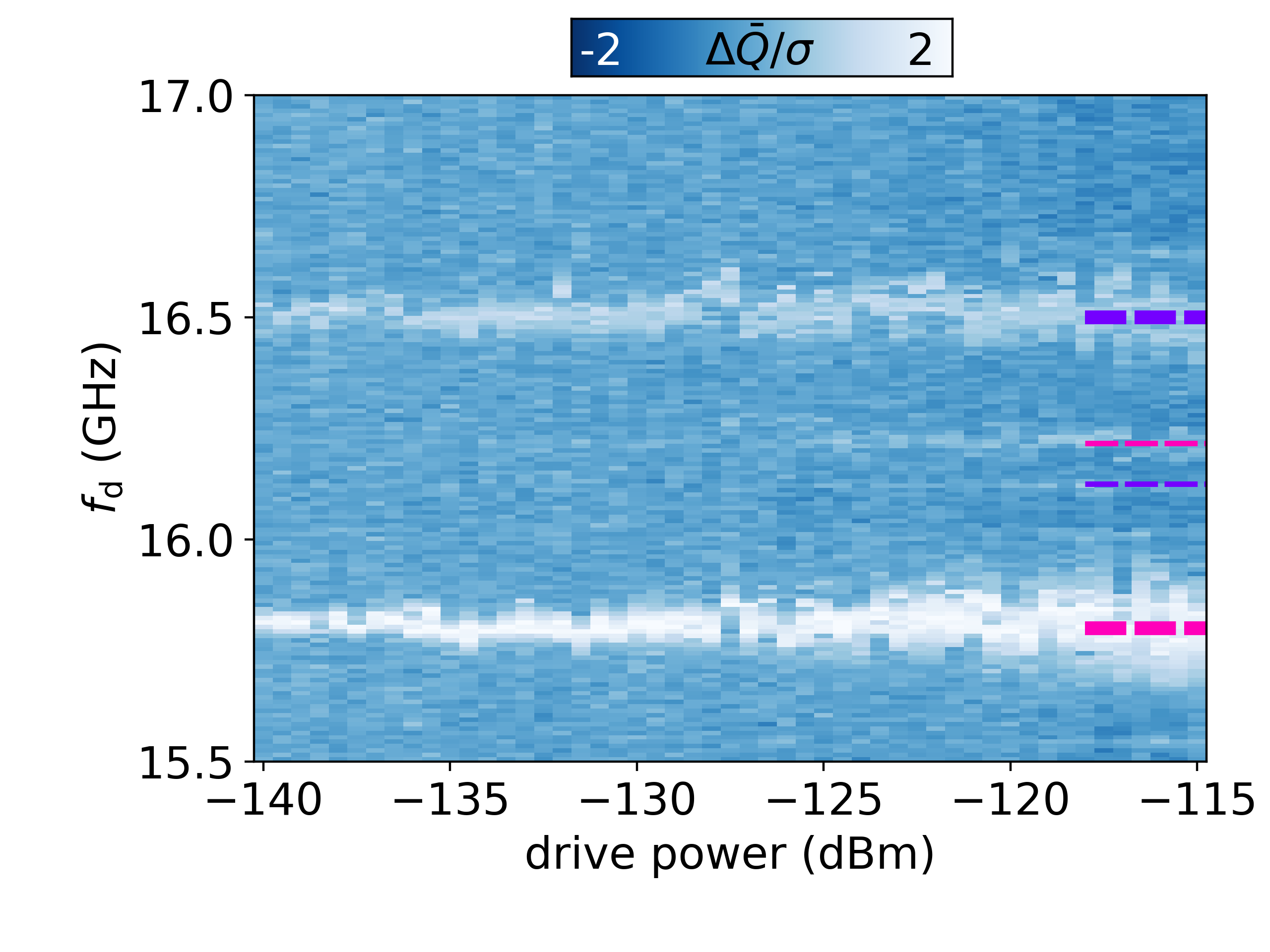} 
	\caption{
    Brightness of the four inter-doublet transitions as a function of estimated drive power at the device. 
    At low powers, only the spin-conserving transitions are visible, but as the power is increased the spin-flipping transitions also appear. 
    Note that the spin-flipping transitions at the maximum power (-115 dBm) are still substantially dimmer than the spin-preserving transitions at the lowest power (-140 dBm). 
    }
\end{figure}

\begin{figure}[H]
	\centering
	\includegraphics[width=3.5in]{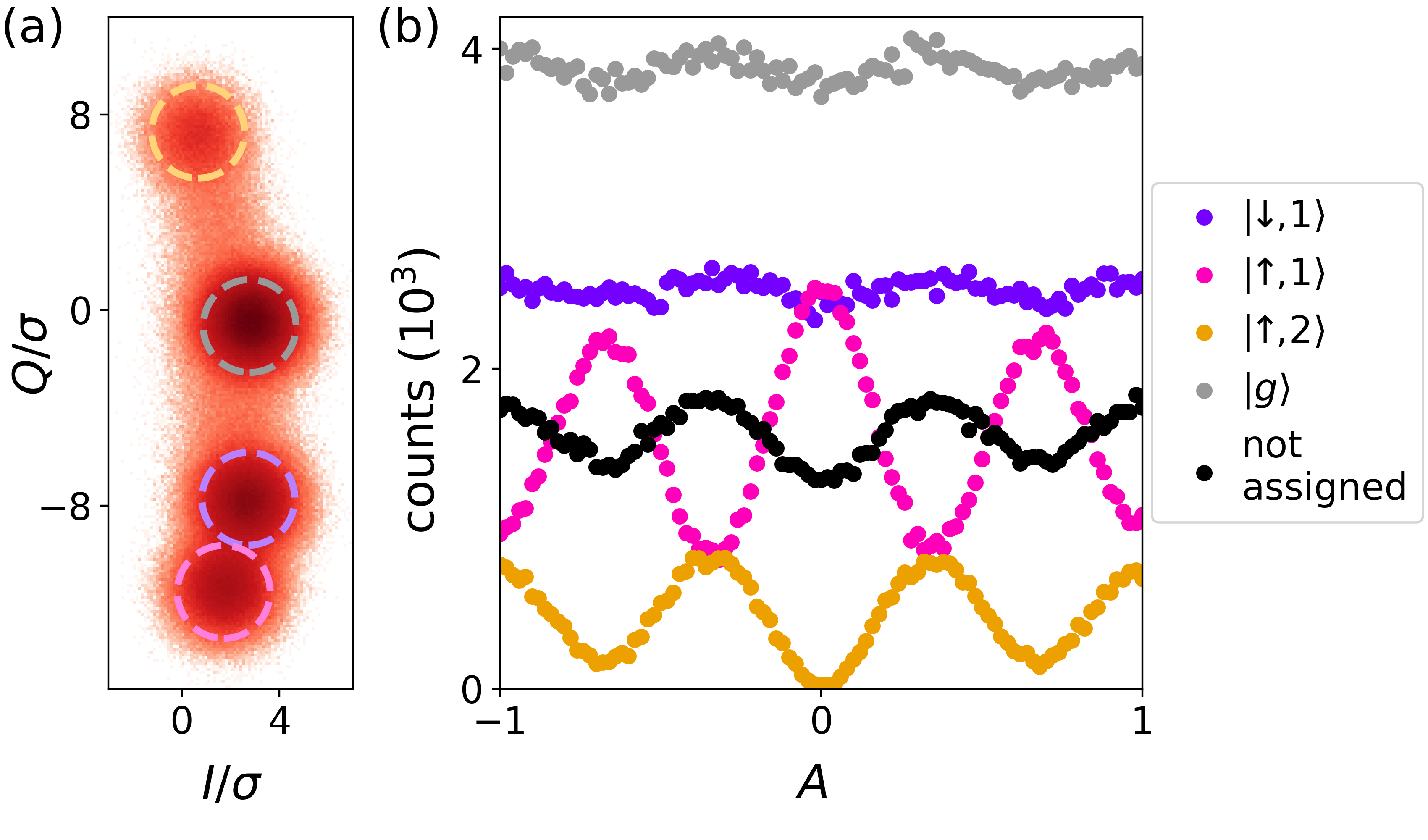} 
	\caption{
	Further detail for the analysis of the Rabi experiment depicted in Fig. 3(d).
	(a) Histogram of all measurement shots taken during the experiment. 
	Shots inside the dashed circles (radius 2$\sigma$) were assigned to the corresponding state. 
	Shots outside these regions were left unassigned. 
	Note that here we also include $\g$ for illustration. 
	(b) At each value of the normalized pulse amplitude $A$, we count the number of points inside each of the four depicted circles in (a).
	The number of unassigned counts is also plotted. 
	See Methods for further details and comments. 
    }
\end{figure}

\begin{figure}[H]
	\centering
	\includegraphics[width=6.5in]{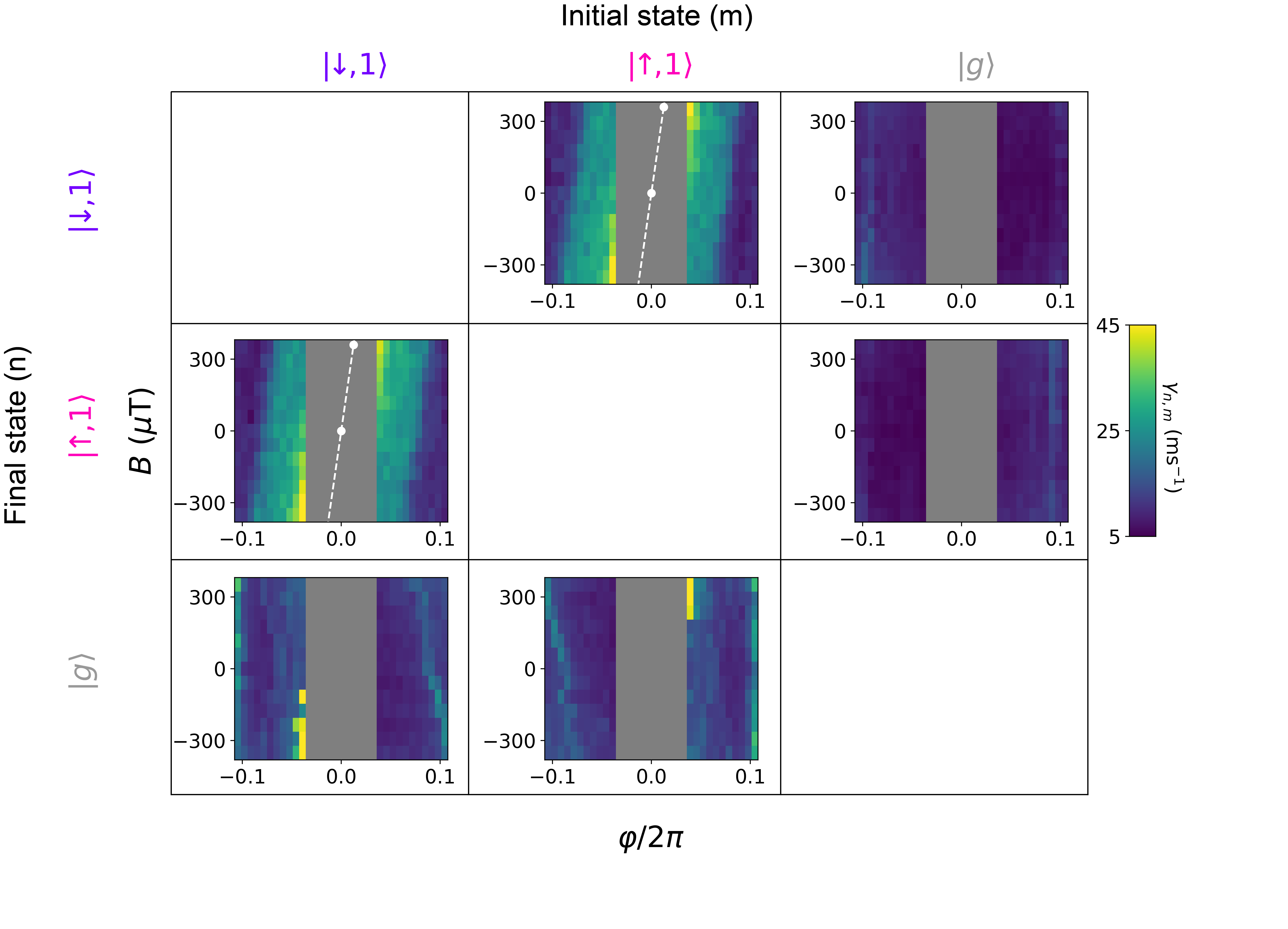} 
	\caption{
	The six extracted transition rates $\gamma_{n, m}$ between $\g$, $\oa$, and $\ob$ as a function of $\varphi$ and $B_\perp$ at the base temperature of the fridge $T = 20~\mK$. 
	White data points on the splin-flipping rate plots indicate the $\oa$/$\ob$ degeneracy point at $B_\perp = 0~\uT, 380~\uT$ (see Main Text Fig. 2(c), Extended Data Fig. 7, Supplementary Information). 
	The dashed lines connect these points and are guides for the eye.   
    }
\end{figure}

\begin{figure}[H]
	\centering
	\includegraphics[width=4in]{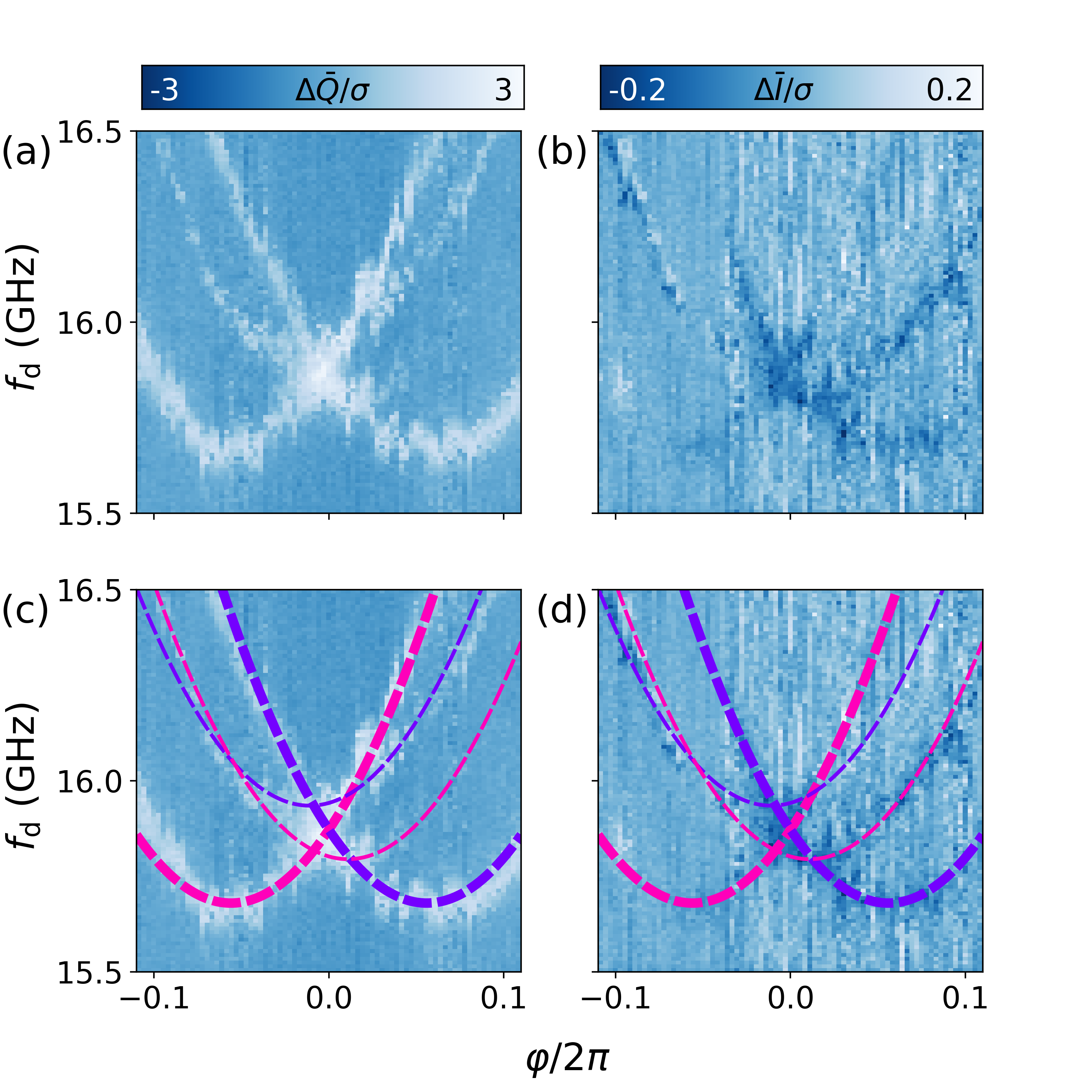} 
	\caption{
    Spectroscopy at $B_\perp$ = 380 $\uT$. 
    Note that here there was a slight overall frequency shift due to a change in the electrostatic environment of the nanowire.
    Here we plot both $I$ (a) and $Q$ (b) to present information in both quadratures.
    The observed instabilities varied with time, and occurred when operating our flux coil at high current. 
    The same data is plotted in (c)/(d), but with overlaid fits. 
    We describe the data by the model used in Fig. 2(c), but we include an additional Zeeman-like term (see Supplementary Information for details). 
    }
\end{figure}

\begin{figure}[H]
	\centering
	\includegraphics[width=6.5in]{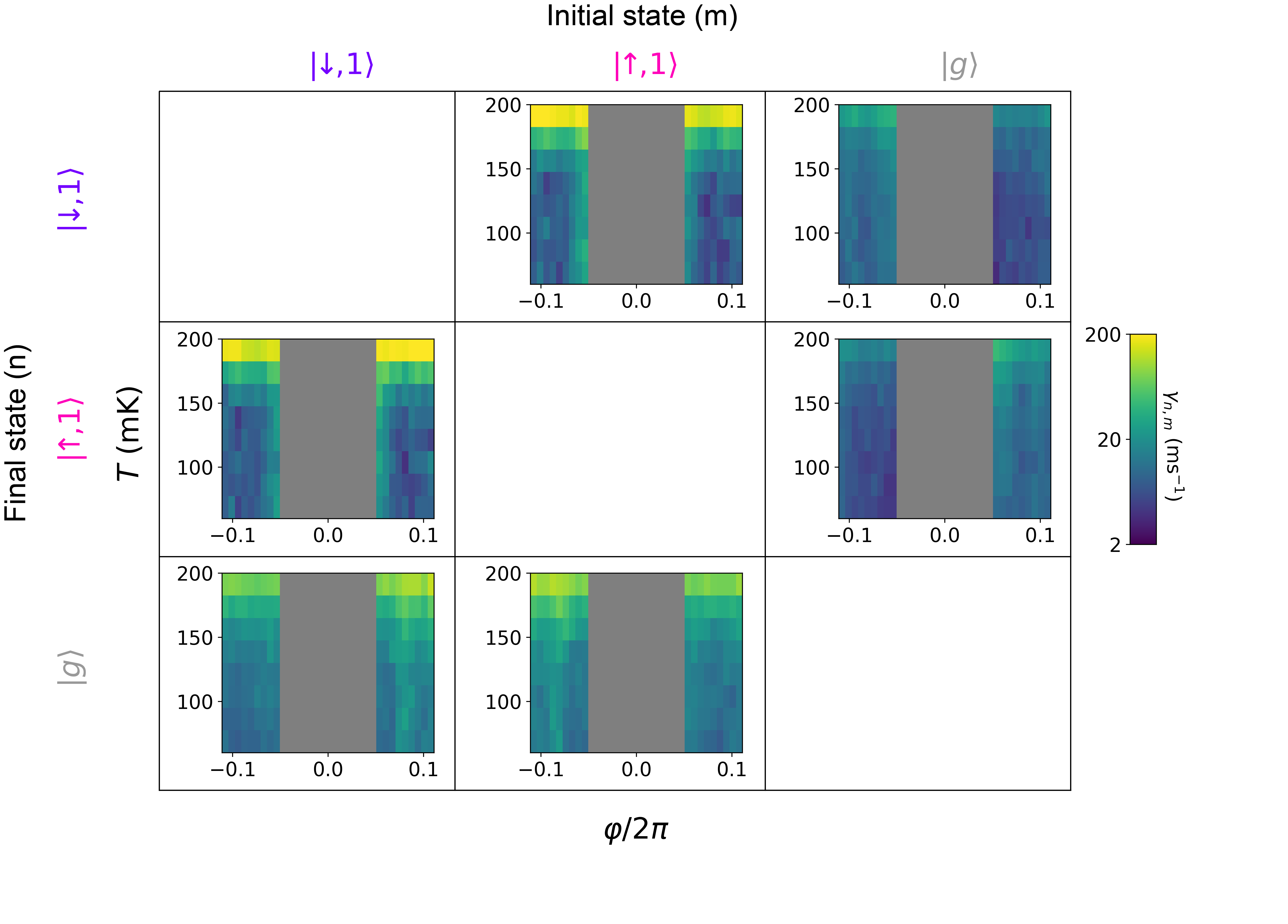} 
	\caption{
	The six extracted transition rates $\gamma_{n, m}$ between $\g$, $\oa$, and $\ob$ as a function of $\varphi$ and mixing chamber temperature $T$ at $B = 0~\uT$.
    }
\end{figure}

\section*{Supplementary Information}

\subsection*{Model of the Andreev levels}
Consider charge carriers of energy $\epsilon \ll \Delta$  moving in a weak link at velocity $\vf$, where $\Delta$ is the pair potential of the superconducting reservoirs. 
As they propagate across the weak link, the charge carriers pick up a phase $\phi_\mathrm{prop} = \epsilon L/\vf \hbar$, where $\ell$ is the length of the weak link. 
When the charge carriers reach the superconducting reservoirs, they undergo Andreev reflection and acquire a phase\supercite{nazarov2009quantum} $\phi_\mathrm{A,\pm} = \epsilon/\Delta  \mp (\varphi + \pi)/2$. 
Here $+(-)$ corresponds to positive (negative) current-carrying charge carriers. 
If the total phase acquired after traveling in the closed loops of Fig. 1(b) is an integer multiple of 2$\pi$, an Andreev level forms. 
Mathematically, this constructive interference condition is expressed as $2 \pi m = 2\phi_\mathrm{prop} + 2\phi_\mathrm{A,\pm}$, where $m \in \mathbb{Z}$. Solving for $\epsilon$ yields Eq. (1) of the main text, repeated here: 

\begin{equation}
   \epsilon(\varphi,s) = \pm  \frac{\Delta \hbar \vf/\ell}{2( \Delta + \hbar \vf/\ell)}(\varphi - \pi (2m + 1))
\end{equation}

\noindent From this, we see that the slope of this dispersion decreases as $\ell$ is increased.
Therefore, the spacing between Andreev levels of different $m$ decreases, much like a Fabry-Perot cavity, quantum dot, or other confined system. 
Spin-orbit interactions can result in a spin-dependent $v_{\mathrm{F},s}$ and the spin-degeneracy of the Andreev levels is broken.
Importantly, whether $v_{\mathrm{F}, \uparrow} > v_{\mathrm{F}, \downarrow}$ or $v_{\mathrm{F}, \uparrow} < v_{\mathrm{F}, \downarrow}$ depends on the sign of the momentum. 
Any physical weak link will have some amount of disorder, which one may model by including point-like spin-conserving scatterers via delta-function potentials\supercite{tosi2019spin}. 
In practice, however, the Andreev spectrum will depend on the exact structure of the weak link disorder.
Nonetheless, a qualitative picture of the nanowire spectrum amounts to linearly-dispersing levels with avoided crossings between levels of like spin [Fig. 1(c)]. 

In this work, we tuned the chemical potential such that a trapped quasiparticle had access to two doublets that we could observe. 
Based on the above discussion, and restricting ourselves to a single quasiparticle excitation, we modeled these doublets beginning with the four linearly-dispersing states $\ket{\uparrow,+}$, $\ket{\uparrow,-}$, $\ket{\downarrow,+}$, and $\ket{\downarrow,-}$. 
To describe these states around $\Phi = 0$, we constructed the phenomenological Hamiltonian

\begin{equation}
    H_\mathrm{A}(\Phi) =  \frac{1}{\Phi_0}
    \begin{bmatrix}
    +m_1(\Phi + \Phi_\mathrm{cross}) & r & 0 & 0 \\
    r & -m_2(\Phi + \Phi_\mathrm{cross}) & 0 & 0 \\
    0 & 0 & +m_2(\Phi - \Phi_\mathrm{cross}) & r \\
    0 & 0 & r & -m_1(\Phi - \Phi_\mathrm{cross}) \\
    \end{bmatrix}
\end{equation}

\noindent Here $m_1 > m_2$ are the slopes of the linearly dispersing Andreev levels, $\Phi_\mathrm{cross}$ is the flux at which the levels cross, and $r$ is a phenomenological parameter that quantifies the strength of the avoided crossing between states of like spin. 
We have ignored an overall offset of the levels within the gap such that the levels cross at zero energy. 
The states $\ket{s,n}$ discussed in the main text correspond to the eigenstates of $H_\mathrm{A}$ with eigenvalues $\epsilon_{s,n}$. 
The choice of spin labels was arbitrary. 
As discussed in the main text, we attribute the four transitions observed in Fig. 2(c) to the inter-doublet transitions of frequency $\omega_{s s', n n'}/2\pi = (\epsilon_{s'_{n'}} - \epsilon_{s,n})/h$. 
Note that these transition frequencies over-constrain the model; i.e. the fourth $\omega_{s s', 12}$ follows when the other three are known. 
The measured values $\omega_{s s', 12}$ thus completely determine the phenomenological parameters that define the Hamiltonian Eq. (2). 
By fitting the spectrum [Fig. 2(c)], we find $m_1 = h\times 22.6~\GHz$, $m_2 = h\times 21.4~\GHz$, $\Phi_\mathrm{cross} = 0.055 \Phi_0$, and $r = h\times 7.6~\GHz$.
Note that $\pm \Phi_\mathrm{cross}$ is also the flux point where the frequencies of the spin-conserving transitions $\omega_{s s', n n'}$ are minimum. 

\subsection*{Model of the nanowire/resonator coupling}

The nanowire and the resonator are coupled because a fraction $p$ of the resonator flux $\Phi_\mathrm{zpf}(a + a^\dagger)$ drops over the nanowire weak link (in the main text, we refer to this flux drop as $\Phi_\mathrm{r} = p \Phi_\mathrm{zpf}$). 
This introduces a perturbation to the nanowire Hamiltonian such that the full system Hamiltonian can be written as\supercite{zazunov2003andreev, bretheau2013localized, Janvier15}
\begin{align}
\begin{split}
    H & = \hbar \omega_\mathrm{r} a^{\dagger}a + H_\mathrm{A} (\Phi + p \Phi_\mathrm{zpf}(a + a^\dagger)) \\
    & \approx \hbar \omega_\mathrm{r} a^{\dagger}a + H_\mathrm{A} (\Phi)+ \frac{d H_\mathrm{A}}{d \Phi} p \Phi_\mathrm{zpf}(a + a^\dagger)+ \frac{1}{2}\frac{d^2 H_\mathrm{A}}{d \Phi^2} (p \Phi_\mathrm{zpf})^2(a + a^\dagger)^2  + ...\\
    & \approx \hbar \omega_\mathrm{r} a^{\dagger}a +H_\mathrm{A} (\Phi)+J p \Phi_\mathrm{zpf}(a + a^\dagger)+ L^{-1} (p \Phi_\mathrm{zpf})^2 a^\dagger a + ...
\end{split}
\end{align}

\noindent where a Taylor expansion is performed in the first step and a rotating-wave approximation in the second\supercite{cristian}. 
We thus find that the resonator flux couples to the current operator $J = \frac{d H_\mathrm{A}}{d \Phi}$ at first order in $p \Phi_\mathrm{zpf}$ and to the inverse inductance operator $L^{-1} = \frac{d^2 H_\mathrm{A}}{d \Phi^2}$ at second order.  
We now consider only the first-order coupling, and compute $J$ from our model $H_\mathrm{A}$: 

\begin{equation}
    J = \frac{d H_\mathrm{A}}{d \Phi}  = 
    \frac{1}{\Phi_0}
    \begin{bmatrix}
    +m_1 & 0 & 0 & 0 \\
    0 & -m_2 & 0 & 0 \\
    0 & 0 & +m_2 & 0 \\
    0 & 0 & 0 & -m_1 \\
    \end{bmatrix}
\end{equation}

\noindent The dispersive shift $\chi_{s,n}$ of $\ket{s,n}$ can then be computed at second order in perturbation theory\supercite{Manucharyan2012}

\begin{equation}
    \chi_{s_n} = -\frac{(p \; \Phi_\mathrm{zpf})^2}{\hbar^2} \sum_{s', n'}\frac{2\omega_{s s', n n'}\left|\bra{s',n'}J\ket{s,n}\right|^2}{\omega_{s s',n n'}^2-\omega_\mathrm{r}^2}
\end{equation}

\noindent where the sum excludes $\{s',n'\} = \{s,n\}$. 
In the bases of Eqns. (2) and (3),  $J$ is diagonal while $H_\mathrm{A}$ is not. 
However, the only off-diagonal elements in $H_\mathrm{A}$ are between states of the same spin. 
As such, $J$ remains block-diagonal in spin when written in the energy eigenbasis (which we do not do explicitly here).
The matrix elements connecting different spins $\bra{\bar{s},n'}J\ket{s,n}$ are thus zero and only the inter-doublet spin-conserving transitions contribute to the dispersive shift:

\begin{equation}
    \chi_{s,n} = -\frac{(p \; \Phi_\mathrm{zpf})^2}{\hbar^2}  \frac{2\omega_{ss,n\bar{n}}\left|\bra{s,\bar{n}}J\ket{s,n}\right|^2}{\omega_{s s,n \bar{n}}^2-\omega_\mathrm{r}^2}
\end{equation}

\noindent This is Eq. (2) of the main text with $f_s = \omega_{s s,n \bar{n}}/2\pi$. 
As discussed in the previous section, the parameters defining $H_\mathrm{A}$ (and therefore $J$) were inferred from the measured spectrum [Fig. 1(c)]. 
This allowed us to calculate the $\Phi$-dependent matrix elements $\bra{s,\bar{n}}J\ket{s,n}$, and therefore the $\Phi$-dependence of $\chi_{s,n}$. 
We found that a value of $p \Phi_\mathrm{zpf}/\Phi_0 = 1.70 \times 10^{-3}$, which is within $10\%$ of an independent calculation (see below), matched the data well in the vicinity of $\Phi_\mathrm{cross}$. 
The coupling strength $g_{\mathrm{c},s} = \frac{p \Phi_\mathrm{zpf}}{\hbar} \frac{}{}\left|\bra{s,\bar{n}}J\ket{s,n}\right|$ is plotted in Extended Data Fig. 2(a). 
As expected, the coupling is peaked around $\Phi_\mathrm{cross}$ where the mixing between the levels is strongest. 

Finally, to translate the predicted $\chi_{s,n}$ to the resonator response, we used the scattering formula for a resonator measured in reflection\supercite{axline2018building}:  

\begin{equation}
    S_{s_n} = \frac{\omega_\mathrm{ro}-(\omega_\mathrm{r}+\chi_{s,n})+i(\kappa_\mathrm{c} -\kappa_\mathrm{i})/2}{\omega_\mathrm{ro}-(\omega_\mathrm{r}+\chi_{s,n})-i(\kappa_\mathrm{c} +\kappa_\mathrm{i})/2}
\end{equation}

\noindent where $\omega_\mathrm{ro} = 2\pi \times 9.18847~\GHz$ is the readout frequency.
After multiplying by a constant complex scale factor to account for the amplitude and phase of our signal $\Gamma_{s_1} = A e^{i \phi} S_{s,1}$, taking the imaginary part gave $Q_{s,1}$ as plotted in Fig. 2(d), and again in Extended Data Fig. 2(b). 
Additionally in Extended Data Fig. 2(b), we plot the expected $\Phi$-dependence of the distributions assuming $g_{\mathrm{c}, s}$ remains constant at the maximum value of $2\pi \times 37.4~\MHz$ (dotted lines). 
In this case, the dispersive shift has much less $\Phi$-dependence than what is measured. 
This illustrates the necessity of using the $\Phi$-dependent $g_{\mathrm{c}, s}$ as calculated from our model of $J$. 

We now return to the inverse inductance $L^{-1}$. 
In the original expansion of Eq. (3), we saw that the resonator and nanowire were coupled at first order in $J$ and second order in $L^{-1}$. 
However, because the current coupling is via the  off-diagonal elements $\bra{s,\bar{n}}J\ket{s,n}$, the dispersive shift [Eq. (6)] is second order in $p \Phi_\mathrm{zpf}$. 
On the other hand, the inverse inductance may induce frequency shifts through its diagonal elements: $\frac{(p \Phi_\mathrm{zpf})^2}{\hbar}\bra{s,n}L^{-1}\ket{s,n}$. 
Thus, both coupling terms contribute frequency shifts at second order in $p \Phi_\mathrm{zpf}$. 
Although within our simple model $L^{-1} = 0$, in reality the full nanowire Hamiltonian is needed to calculate both $J$ and $L^{-1}$. 
Such calculations have been performed in the ``short junction'' regime ($\hbar \vf/L \gg \Delta$) for pair transitions\supercite{zazunov2003andreev, bretheau2013localized}, and for intra-doublet transitions in longer weak links\supercite{park2017andreev}, but it is an unsolved theoretical problem for the inter-doublet transitions explored in the current work.
It is worth noting that in the short junction case, the value of the inverse inductance computed from $\bra{s,n}L^{-1}\ket{s,n}$ can be much smaller than what one might expect from the $\Phi$-dispersion of the energies $\frac{d^2\epsilon_{s,n}}{d \Phi^2}$. 
Moreover, while we anticipate that the dispersive shift due to $L^{-1}$ around $\Phi = 0$ should be negative, the shift around $\Phi = -0.5 \Phi_0$ should be positive and of similar magnitude.
In contrast, the shifts due to the current coupling should always be negative so along as $\omega_{s s,n \bar{n}} > \omega_r$ (see Eq. (6)). 
We only observe negative frequency shifts over the entire $\Phi$ range [Extended Data Fig. (3)], which we interpret as the current coupling being dominant. 
The remaining discrepancies between our simplified model and the data for the dispersive shift are beyond the scope of this work and may involve additional subtleties in the Andreev Hamiltonian not captured here (see Ref. \cite{zazunov2003andreev} and Appendix B of Ref. \cite{bretheau2013localized}).

We now outline the calculation of $p \Phi_\mathrm{zpf}$ that was performed during the construction of the experiment. 
We modeled the resonator as a $\sim \lambda/4$ length of transmission line with impedance $Z_0 = 70~\Omega$.
This impedance was larger than its purely geometrical value due to the sheet inductance $0.6~\pH/\mathrm{square}$ of the NbTiN. 
The coupling capacitance [Extended Data Fig. 1(c)] to the readout transmission was approximated as an open boundary condition. 
The other boundary condition was set by the nanowire loop [Extended Data Fig. 1(d)], which was modeled as an inductance $L$. 
Because the nanowire inductance was larger than the shared inductance, we neglected the nanowire inductance and calculated $L = 68~\pH$ based on both the geometric and kinetic inductance of the shared trace. 
Using standard microwave formulas (see ref. \cite{pozar2009microwave}, Eq. 2.44), we calculated the fraction of the mode voltage (which is the same as the fraction of the mode flux) that dropped over the shared inductance to be $p = 0.057$. 
The zero-point fluctuations of the resonator were calculated as $\Phi_\mathrm{zpf} = \sqrt{\hbar Z_\mathrm{res}/2}$, where the resonator impedance is related to the transmission line impedance by  $Z_\mathrm{res} = 4 Z_0/\pi$ (see ref. \cite{pozar2009microwave}, Eq. 6.30b/c). 

\subsection*{Purcell limit of the inter-doublet decay rate}


Due to the coupling between the Andreev levels and the resonator, a quasiparticle occupying an Andreev level can lose energy through the resonator (Purcell effect\supercite{Pur46}).
At the bias point for the data displayed in Fig. 3, the Purcell-induced energy decay rate for a quasiparticle in the higher doublet is $\gamma_\mathrm{P} \approx (\kappa_\mathrm{c} + \kappa_\mathrm{i})\frac{g_\mathrm{c}^2}{(\omega_{\uparrow \uparrow, 12} - 2 \pi \fr)^2} = \frac{1}{4.3~\ms}$. 
This is roughly three orders of magnitude higher than the observed decay rate. 
\subsection*{Dependence of transition rates on temperature and magnetic field}

As summarized in the Methods section, we used a hidden Markov model to extract the transition rates $\gamma_{n, m}$ between $\g$, $\oa$, and $\ob$, where $m$ labels the initial state and $n$ the final state. 
We investigated these rates as a function of the phase across the nanowire weak link $\varphi$, a magnetic field applied approximately perpendicular to the device substrate $B_\perp$ [Extended Data Fig. 6], and the temperature of the mixing chamber $T$ [Extended Data Fig. 8]. 

As discussed in the main text, we found strong dependence of the spin-flip rates on $\varphi$ correlated with the energy splitting between $\oa$ and $\ob$. 
Additionally, we note that the $B_\perp$-dependence is consistent with a Zeeman-like shift of the Andreev levels.
We checked this interpretation by measuring the $\varphi$-dependence of the nanowire transition spectrum at $B_\perp = 380~\uT$ [Extended Data Fig. 7].
We modeled the spectrum using the same approach as for Fig. 2(c),  but with an additional Zeeman-like term $\epsilon_{\downarrow,n} \rightarrow \epsilon_{\downarrow,n} - E_\mathrm{Z}$, $\epsilon_{\uparrow,n} \rightarrow \epsilon_{\uparrow,n} + E_\mathrm{Z}$.
Note that within this model, only the spin-flipping transitions are affected by $E_\mathrm{Z}$. 
We found that $E_\mathrm{Z} \approx h \times 35~\MHz$ qualitatively described the data, which corresponds to a shift in the $\oa$/$\ob$ degeneracy point to $\varphi = 2 \pi \times 0.013$, consistent with the $B_\perp/\varphi$ slope observed in the spin-flipping rates (white dashed lines in the $\gamma_{s,\bar{s}}$ plots of Extended Data Fig. 6).  
However, a systematic study of $E_\mathrm{Z}$ versus $B_\perp$ was made difficult by instabilities induced by the large coil current necessary to generate $B_\perp$ [Extended Data Fig. 7]. 

The two rates corresponding to quasiparticles entering the nanowire weak link were almost entirely unaffected by both $\varphi$ and $B_\perp$.
Curiously, the rates corresponding to the inverse process (quasiparticles leaving the nanowire weak link) were generally higher and exhibited some weak features. 
In particular, the $\varphi$-dependence of $\gamma_{0, {s1}}$ at $B_\perp = 0~\uT$ exhibits a peak for the higher energy spin state [Extended Data Fig. 6].
Applying a positive (negative) $B_\perp$ resulted in a negative (positive) shift of the $\varphi$-dependence, opposite that of the spin-flip rates.
This is consistent with the higher-energy $\ket{s,1}$ coming into resonance with a cold mode through which the quasiparticle can be evacuated.  

We also investigated the dependence of the rates on the temperature of the mixing chamber $T$ [Extended Data Fig. 8].
Surprisingly, we observed the spin-flipping rates were unaffected by increasing $T$ until $\sim 150~\mK$. 
Moreover, the temperature dependence of the spin-flip rates was purely additive to the low-temperature behavior and did not itself depend on $\varphi$. 
This suggests that the mechanism resulting in the low-temperature $\varphi$-dependence is not the same as the mechanism that kicks in at higher temperatures. 
Similarly, the quasiparticle-switching rates were unaffected by increasing $T$ until $\sim 150~\mK$.
These rates should be related to the fraction of broken Cooper pairs in the circuit, which has been shown in other contexts to be temperature independent below a similar temperature scale due to non-equilibrium quasiparticles present at low temperatures\supercite{martinis2009energy, barends2011minimizing, catelani2011relaxation, serniak2018hot}. 
Thus, this data is consistent with the known phenomenology of non-equilibrium quasiparticles in superconducting circuits. 

\printbibliography[keyword={supplement},title={References (Supplement)}]
 
\end{document}